\documentclass[preprint2]{aastex} 

\shorttitle{The influence of Host Galaxies in Type Ia Supernova Cosmology}
\shortauthors{Uddin et al.}

\usepackage{amsmath}
\usepackage{enumitem}
\usepackage{threeparttable}
\usepackage{adjustbox}

\usepackage{lscape}
\usepackage{longtable}

\usepackage{natbib}
\usepackage{hyperref} 

\hypersetup{colorlinks,citecolor=blue,linkcolor=blue,urlcolor=blue}

\begin{document}




\title{The influence of Host Galaxies in Type Ia Supernova Cosmology}


\author{Syed A. Uddin\altaffilmark{1,2,3}, Jeremy Mould\altaffilmark{2,3}, Chris Lidman\altaffilmark{3,4}, Vanina Ruhlmann-Kleider$^5$, and Bonnie R. Zhang\altaffilmark{3,6}}
\affil{$^1$Purple Mountain Observatory, Chinese Academy of Sciences, Nanjing, Jiangshu, China}
\affil{$^2$Centre for Astrophysics and Supercomputing, Swinburne University of Technology, Melbourne, VIC, Australia}
\affil{$^3$ARC Centre of Excellence for All-Sky Astrophysics (CAASTRO)}
\affil{$^4$Australian Astronomical Observatory, North Ryde, NSW, Australia}
\affil{$^5$CEA, Centre de Saclay, Irfu/SPP, 91191 Gif-sur-Yvette, Paris, France}
\affil{$^6$Research School of Astronomy and Astrophysics, Australian National University, Canberra, ACT 2611, Australia}

\email{saushuvo@gmail.com}




\begin{abstract}

We use a sample of 1338 spectroscopically confirmed and photometrically classified Type Ia Supernovae (SNe Ia), sourced from the CSP, CfA, SDSS-II, and SNLS supernova samples, to examine the relationships between SNe Ia and the galaxies that host them. Our results provide confirmation with improved statistical significance that SNe Ia, after standardization, are on average more luminous in massive hosts (significance $\rm > 5 \sigma$), and decline more rapidly in massive hosts (significance $\rm > 9\sigma$) and in hosts with low specific star formation rates (significance $\rm > 8\sigma$). We study the variation of these relationships with redshift and detect no evolution.  We split SNe Ia into pairs of subsets that are based on the properties of the hosts, and fit cosmological models to each subset. Including both systematic and statistical uncertainties, we do not find any significant shift in the best-fit cosmological parameters between the subsets. Among different SN Ia subsets, we find that SNe Ia in hosts with high specific star formation rates have the least intrinsic scatter ($\rm \sigma_{int}=0.08\pm0.01$) in luminosity after standardization.


\end{abstract}





\section{Introduction}

The standardization of SN Ia luminosities (\citealt{phillips93}, \citealt{tripp98}) and their subsequent use as distance indicators led to the surprising discovery that the expansion of the universe is accelerating (\citealt{riess98}, \citealt{perlmutter99}). Present observations (\citealt{planck15}) are consistent with the idea that the current epoch of acceleration is being driven by a cosmological constant (\citealt{weinberg89}). At present, systematic uncertainties in SN Ia cosmological measurements are comparable with statistical uncertainties (\citealt{betoule14}; hereafter B14). Currently, the largest source of systematic uncertainty comes from calibration. It is not unreasonable to suppose that, as calibrations improve, the astrophysics of SNe Ia may become an important source of systematic uncertainty. 

It is well established that there are correlations between the brightness of SNe Ia and the galaxies that host them (e.g., \citealt{sullivan10}, \citealt{childress13} etc. and discussed below). Over the range of redshifts where SNe Ia are used in cosmology, the galaxy population evolves significantly. At higher redshifts, galaxies were, on average, younger, less massive, and were forming many more stars. As the average properties of galaxies evolve with redshift, so will the average properties of the SN Ia population, and this, in combinations with the magnitude-limited selection of SNe Ia, may lead to biases in cosmological parameters.

A number of studies show the existence of correlations between the properties of SNe Ia and the properties of galaxies that host them. Some studies (e.g., \citealt{neill09}, \citealt{kelly10}, \citealt{lampeitl10a}, \citealt{sullivan10}) used multiband photometry, while others (e.g., \citealt{gallagher08}, \citealt{dandrea11}) used spectra to derive host galaxy properties. Some used both (e.g., \citealt{childress13}, \citealt{pan14}, \citealt{campbell16}, \citealt{wolf16}). With variable significance, most studies show similar results, namely that SNe Ia are brighter (after correcting for color and light-curve width) when hosted in high-mass and low-sSFR\footnote{specific Star Formation Rate} galaxies, that SNe Ia have narrower stretch (or are faster declining) when hosted in high-mass, low-sSFR galaxies, and that redder SNe Ia appear in massive, older, and higher metallicity hosts. \cite{galbany12}, using SDSS data, found that the color of SNe Ia in spiral galaxies become bluer as they approach the centers of their hosts. 

Other studies have used the properties of the host to select subsamples of SNe Ia that have smaller Hubble residuals.
Splitting host galaxies into different morphological types, \cite{sullivan03} found a smaller scatter in the Hubble diagram with SNe Ia in early-type galaxies. \cite{hill16} recently showed with lower significance ($1.4\sigma$), that SNe Ia exploding in the outskirts of galaxies have smaller Hubble residuals.

Using integral field spectroscopy data, \cite{rigault13} discovered that SNe Ia in hosts that are locally emitting $H\alpha$ are more homogenous. Recently \cite{kelly15} found that SNe Ia in hosts with high ultraviolet surface brightness can be calibrated to yield precise ($\sim 3-4 \%$) distance measurements. Previously, \cite{hayden13} showed that metallicity can be used to reduce the Hubble residual.

It is not clear what drives these correlations. Dust, progenitor age, progenitor metallicity  or a combination of all three could be responsible. Recent work (\citealt{childress14}) qualitatively showed that the observed trends in the Hubble residual offset with host mass may come from progenitor age differences. The evolution of these offsets between host types with redshift, not known conclusively, may introduce additional complexity.

Cosmological parameters seem not to differ when they are derived from SNe Ia in different host environments. For example, \cite{sullivan11} derived cosmological parameters with SNe Ia in low-mass and high-mass hosts and find no significant difference in the derived cosmological parameters. Similarly, \cite{campbell16} also did not find any significant shifts in cosmological parameters when host galaxy properties are used as a correction term in deriving SN Ia luminosities.

Studying the impact  these correlations have on SN Ia cosmology is becoming increasingly important as SN Ia surveys become larger, such as DES$\footnote{http://darkenergysurvey.org}$ and  LSST$\footnote{http://www.lsst.org}$, where many thousands of SNe Ia will be discovered over a broad redshift range. For example, DES will mostly use redshifts from SN Ia host galaxies when constructing SN Ia Hubble diagram. These redshifts will come from large spectroscopic campaigns such as OzDES$\footnote{http://www.mso.anu.edu.au/ozdes/}$, where it is easier to obtain redshifts from massive, star-forming galaxies \citep{yuan15}. A DES SN Ia sample will therefore be biased in terms of host properties. 


 Host galaxy properties from different surveys are derived in different ways. When compiling such a sample from different surveys to study SNe Ia properties in their host galaxies, it is desirable to have host galaxy properties derived in a uniform manner. In this paper we build a sample of literature SNe Ia along with their host galaxy properties derived consistently from $ugriz$ photometry. Our study is the first one to use such a large sample of SNe Ia to study the relations between the properties of SN Ia and their hosts.  With this uniform sample we study how SN Ia stretch ($x_1$), color ($c$), and Hubble residual ($\Delta \mu$) vary with host galaxy properties such as stellar mass, sSFR, and galaxy shape. We also study how the properties of SNe Ia depend where they explode in their hosts. For a given property of the host (e.g., mass), we split the sample into two subsets (e.g., low-mass and high-mass). We then study the properties of SNe Ia in these subsets and search for differences.  We also study the redshift evolution of these differences. We fit cosmological models within $\Lambda$CDM and $w$CDM to each subset and look at the shifts in best-fit cosmological parameters. Throughout this paper we have assumed the universe is flat.

The outline of this paper is as follows. In Section \ref{sample}, we describe our sample. In Section \ref{correlations}, we describe our results on the offsets of SN Ia properties between subsets, the redshift evolution of these offsets, and cosmological fits to these subsets. We compare our results with other published studies in Section \ref{dis} along with other topic. We conclude by summarizing in Section \ref{summary} and discuss future prospects.

\section{Sample}\label{sample}

\subsection{Supernova Surveys}\label{survey}

In this paper, we combine two different compilations of SNe Ia. In one compilation, all the SNe Ia are spectroscopically confirmed, and in the other compilation, all the SNe Ia are photometrically classified with no real-time spectroscopic confirmation. Redshifts for the later compilation come from spectra of their host galaxies. The SN Ia samples in this paper come from the following surveys:

\textbf{SuperNova Legacy Survey (SNLS):} SNLS pioneered the rolling survey, where the same patch of the sky is repeatedly imaged to discover new supernovae, while at the same time obtaining light curves of previously discovered SNe. The SNLS survey was conducted between 2002 and 2008 with the 3.6 meter Canada-France-Hawaii Telescope (CFHT) located atop the summit of Mauna Kea on the island of Hawai'i. The instrument MegaCam \citep{boulade03} was used to image four one-square degree fields located in areas of low galactic extinction in the $g_M$, $r_M$, $i_M$, and $z_M$ bands. Each field was visited four or five times during each lunar cycle. Images were processed quickly to detect live transients using two different pipelines \citep{perrett10}. In the redshift range $0.2<z<1.1$, SNLS discovered $\sim 1000$ supernovae from 5 years of data (B14), of which 427 are spectroscopically confirmed to be of type SN Ia (Balland et al. in preparation). For spectroscopic confirmation, several 8-10 meter telescopes were used, including the Gemini North and South telescopes, the Very Large Telescope, and the Keck telescopes. SN spectra have been published in several papers, including \cite{howell05}, \cite{bronder08}, \cite{ellis08}, \cite{balland09}, and \cite{walker11}. Analysis and classification of SN spectra are described in \cite{howell05} , \cite{balland09} and (Balland et al. in preparation).  Of the spectroscopic sample, 242 SNe Ia were used for the 3-year cosmological analysis in \cite{conley11} and in \cite{sullivan11} in contrast to the first year cosmological analysis, in which 71 SNe Ia were used \citep{astier06}. Using the same 3-year sample, studies of the host galaxies were undertaken in \cite{sullivan10} and in \cite{sullivan11}.

On the other hand, a deferred analysis of the first three years of SNLS data has also been conducted independently of the real-time pipelines. This analysis led to the definition of a sample of 485 photometrically identified SNe Ia, as presented in \cite{bazin11}. In this sample, 246 events are also part of the spectroscopic sample. Among the 239 remaining events, host spectroscopic redshifts have been obtained for 92 events in dedicated observations at the AAOmega spectrograph (\citealt{lidman13}). These 92 events make the subsample of the SNLS photometric sample used in this paper.


\textbf{Sloan Digital Sky Survey (SDSS) - II Supernova Survey:}  The SDSS-II supernova survey used the 2.5 meter SDSS telescope \citep{gunn06} located at the Apache Point Observatory in New Mexico, USA. In this program, a part of the sky called stripe 82 was repeatedly scanned in the $ugriz$ bands for supernovae during the northern falls of 2005, 2006, and 2007. Stripe 82 is a $2.5^{\circ}$ wide area along the celestial equator between right ascensions of 20h and 04h. Each observation consisted of a scan with an equivalent exposure time of 55 seconds simultaneously in the $ugriz$ bands. Live transients were identified using image-subtraction. Scene modeling photometry \citep{holtzman08} was used to derive light-curves after the transients were discovered.

Visual inspection and model fitting of the light-curves were performed to select SN Ia candidates, which were then spectroscopically confirmed using a number of telescopes \citep{frieman08}. Spectra were analysed to determine SN types as described in \cite{zheng08}. SDSS-II has released the final version of the light-curves of all transients from the full 3-year survey in \cite{sako14}. Among them are $\sim$500 spectroscopically confirmed SNe Ia and $\sim$900 photometrically identified SNe Ia with host spectroscopic redshifts. SDSS-II supernovae are in the intermediate redshift range with $z<0.5$. Using the spectroscopically confirmed SNe Ia, the first year cosmological results were shown in \cite{lampeitl10b}. A number of papers were also dedicated to analyzing the host galaxies, such as \cite{lampeitl10a}, \cite{dandrea11}, \cite{gupta11}, \cite{campbell16}, and \cite{wolf16}. SDSS-II has used photometrically identified SNe Ia \citep{campbell13} to place constraints on cosmological parameters.

The most recent analysis presenting cosmological constraints that include SNe Ia from SNLS and SDSS is the Joint Light-curve Analysis (JLA; B14). An important aspect of this analysis is that the photometry of both SNLS and SDSS-II SNe Ia have been recalibrated with respect to the HST CALSPEC standards \citep{bohlin04}. Light-curve fitting criteria to create a sample for cosmological study are described in \cite{guy10}.
 
\textbf{Carnegie Supernova Project (CSP):} The CSP was conducted between 2004 and 2009 \citep{hamuy06} at the Las Campanas Observatory in Chile using several telescopes, most notably the 1 meter Swope and 2.5 meter du Pont telescopes. Unlike the rolling surveys (e.g. SNLS and SDSS-II), the supernovae observed by CSP were first discovered in surveys that searched nearby galaxies, such as the Lick Observatory Supernova Search \citep{li00}. CSP has obtained light-curves in the $u'gir'i'BVYJHK_s$ bands, which range from the near-ultraviolet to the near-infrared. The photometric data were released in two stages: DR1 \citep{contreras10} and DR2 \citep{stritzinger11} and consist of $\sim 100$ SNe Ia within $z<0.07$. CSP has also developed their own light-curve fitting software called SNooPy (see Sec.~\ref{lc}).  

\textbf{Centre for Astrophysics (CfA) Supernova Survey:} Astronomers at the Harvard-Smithsonian CfA have been collecting photometric and spectroscopic data of local SNe Ia since 1993 \citep{riess99}.  The CfA supernova program uses the 1.2 meter telescope at the Fred Lawrence Whipple observatory using the $UBVIr'i'$ bands. They have four successive data releases: CfA1 (22 SNe Ia, \citealt{riess99}), CfA2 (44 SNe Ia, \citealt{jha07}), CfA3 (185 SNe Ia, \citealt{hicken09}), and CfA4 (94 SNe Ia, \citealt{hicken12}). SNe Ia from CfA3 were extensively used to study host galaxies in \cite{kelly10} and in \cite{neill09}. 

SNe Ia from various surveys that are used in this paper are summarized in Table \ref{snspec} and Table \ref{snphoto}.

\subsection{SN Ia Light-Curve Fitting}\label{lc}
SN surveys produce light-curves (photometric points at different epochs) that we use to derive light-curve parameters. \citet{guy10} studied in detail the sampling requirements needed to obtain reliable estimates of the light-curve parameters. Following that paper and defining SN Ia phase as $\tau=(T_{obs}-T_{max})/(1+z)$, where $T_{obs}$ is the epoch of observation and $T_{max}$ is the epoch of the maximum brightness in the $B$-band, these requirements are:

\begin{itemize}
\item [1. ]Four measurements or more within $-10<\tau<+35$ days
\item [2. ]At least one measurement in the range $-10<\tau<+5$ days and one in the range $+5<\tau<+20$ days
\item[3. ]One measurement or more in at least in two filters in the range $-8<\tau<+10$ days
 
\end{itemize}

We use SNe Ia that pass these sampling criteria and fit their light-curves with a light-curve fitter to determine SN Ia properties such as light-curve width or stretch\footnote{Throughout the paper we will use the term stretch.}, color, and peak $B$-band magnitude. Stretch and color are necessary for correcting the observed peak $B$-band magnitudes. They are also important for the photometric selection described in Sec.~\ref{photosel}. 

There are a handful light-curve fitters available. They are: the Spectral Adaptive Light-curve Template (SALT, \citealt{guy07}), the Multicolor Light Curve Shape (MLCS, \citealt{riess97}), and the SuperNovae in Object Oriented Python (SNooPy). In this paper, we use SALT version 2.4 (hereafter SALT2.4\footnote{http://supernovae.in2p3.fr/salt/doku.php}) as it is the most widely used and tested light-curve fitter in the literature. Unlike some other light-curve fitters, SALT2.4 does not separate intrinsic SN Ia color from reddening by dust in the host. SALT2.4 provides a number of executables to derive properties of SNe Ia. For example, $snfit$ is used to derive stretch, color and peak $B$-band magnitudes and their uncertainties. The other useful executable is $snmag$, which calculates rest-frame $griz$-band magnitudes at the time of $B$-band peak brightness. We use these magnitudes to perform color-magnitude selection of photometrically classified SNe Ia as described next.

%
%

\subsection{SN Ia Selection Criteria}

\subsubsection{Spectroscopically Confirmed Sample}

Table $\ref{snspec}$ lists the number of spectroscopically confirmed SNe Ia from the surveys that are used in this paper. The numbers here are smaller than the number of SNe Ia discovered in these surveys because we restrict our study to SNe Ia for which host galaxies are clearly identified. Some studies (e.g., \citealt{sullivan10}) have used apparently host-less SNe Ia by placing an upper limit of host mass. This essential criteria only affects SNe Ia from SDSS-II and SNLS. Using this criteria, we lose $18\%$ of the SNe Ia from SNLS, and $15\%$ of the SNe Ia from SDSS-II. Host galaxies are always identified for SNe Ia from CSP and CfA as these surveys followed SNe Ia that were discovered in nearby galaxies. For SDSS-II and SNLS, SN Ia light-curves come from JLA (B14). For CSP, CfA3, and CfA4, we have obtained light-curves from the corresponding sources listed in Table \ref{snspec}. The total number of spectroscopically confirmed SNe Ia used in this paper is 583.

\begin{table}[htp]
\caption{Spectroscopically confirmed SN Ia sample used in this paper.}
\begin{center}
\begin{tabular}{lcl}
\hline
Source & Number & Reference\\
\hline
CSP &27 & \cite{stritzinger11}\\
CfA3 & 32 &  \cite{hicken09}\\
CfA4 & 17 &  \cite{hicken12}\\
SDSS & 311 &  \cite{sako14}\\
SNLS & 196 &  \cite{betoule14}\\
\hline
Total & 583\\
\hline
\end{tabular}
\end{center}
\label{snspec}
\end{table}

Spectroscopically confirmed SNe Ia are less likely to have been misclassified compared to their photometrically identified counterparts, since the classification is made from spectral features. 

%


\subsubsection{Photometrically Classified Sample}\label{photosel}

Table \ref{snphoto} lists the number of photometrically classified SNe Ia from each survey that are used in this paper.  In this case, since their identification relies on their light-curves and host spectroscopic redshifts, the contamination from other transients will be higher. In order to  minimize the contamination without loosing too many real SNe Ia, the photometric selection of SNe Ia needs a few extra steps beyond the light-curve sampling. 

\begin{table}[htp]
\caption{Photometrically classified SN Ia sample used in this paper. Redshifts for these SNe Ia come from there host galaxy. SNe Ia listed in this table do not appear in the spectroscopic sample.}
\begin{center}
\begin{tabular}{lcl}
\hline
Source & Number & Reference\\
\hline
SDSS & 661 &  \cite{sako14}\\
SNLS & 94 &  \cite{bazin11}\\
\hline
Total & 755\\
\hline
\end{tabular}
\end{center}
\label{snphoto}
\end{table}

In this paper, photometrically classified SN Ia samples come from SDSS-II (full survey) and SNLS (3-year survey). We summarize the SN Ia selection process for these two surveys.

Transient objects from the SDSS-II supernova survey are first identified using the SDSS-II difference imaging pipeline \citep{sako08}. Transients with positive flux, 3 sigma above the noise level, in at least two contiguous pixels, are selected for further analysis. SDSS-II uses the PSNID (Photometric SuperNova IDentification) software to identify different SN types from the identified transients \citep{sako11}. The software calculates Bayesian probabilities of a transient being of Type Ia, Ib/c, or Type-II SN by comparing observed photometry against a grid of SN Ia and core-collapse templates. An extension of PSNID adopts a kd-tree Nearest-Neighbour algorithm (PSNID/NN) to further refine the classification. Using PSNID as an initial fitter and PSNID/NN as the secondary method, the SDSS-II photometric selection has the following criteria:

\begin{itemize}
\item[1.] $P_{Ia}>P_{Ibc}$ and $P_{Ia}>P_{II}$
\item[2.] $P_{NN,Ia}>P_{NN,Ibc}$ and $P_{NN,Ia}>P_{NN,II}$
\item[3.] $P_{fit}>0.01$
\end{itemize}

where $P$ is the Bayesian probability. The light-curve sampling has the requirements of at least two observations between $-5\le \tau \le +5$ days and $+5\le \tau \le +15$ days, where $\tau$ has been defined previously in Sec.~\ref{lc}.

 SNLS uses SALT (version 2.0) to produce synthetic light-curves needed to compare with observed light-curves. SN-like events are selected using four steps. These are \citep{bazin11} :

\begin{itemize}
\item[1.] Searching for a significant flux variation
\item[2.] Checking a SN-like variation in multiple filters
\item[3.] At least one pre-max and one post-max data point in both $i_M$ and $r_M$ within $-30<T_{max}<+60$, where $T_{max}$ is the epoch of peak brightness
\end{itemize}

After selecting SN-like events SNLS performs further selection to find SN Ia. These criteria are:

\begin{itemize}
\item[1.]  At least one measurement in the range $-10<\tau<+5$
\item[2.] At least one measurement  in the range $+5<\tau<+20$
\item[3.] At least one color measurement in the range $-10<\tau<+35$ from $(g-r)$, $(r-z)$, or $(i-z)$
\item[4.] Reject SNe Ia with poor fits. Requirements are $\chi_{\nu}^2<10$ of the $g$-band and $\chi_{\nu}^2<8$ for $r$, $i$, and $z$
\end{itemize}

\begin{figure}[htbp]
\begin{center}
\includegraphics[width=\columnwidth]{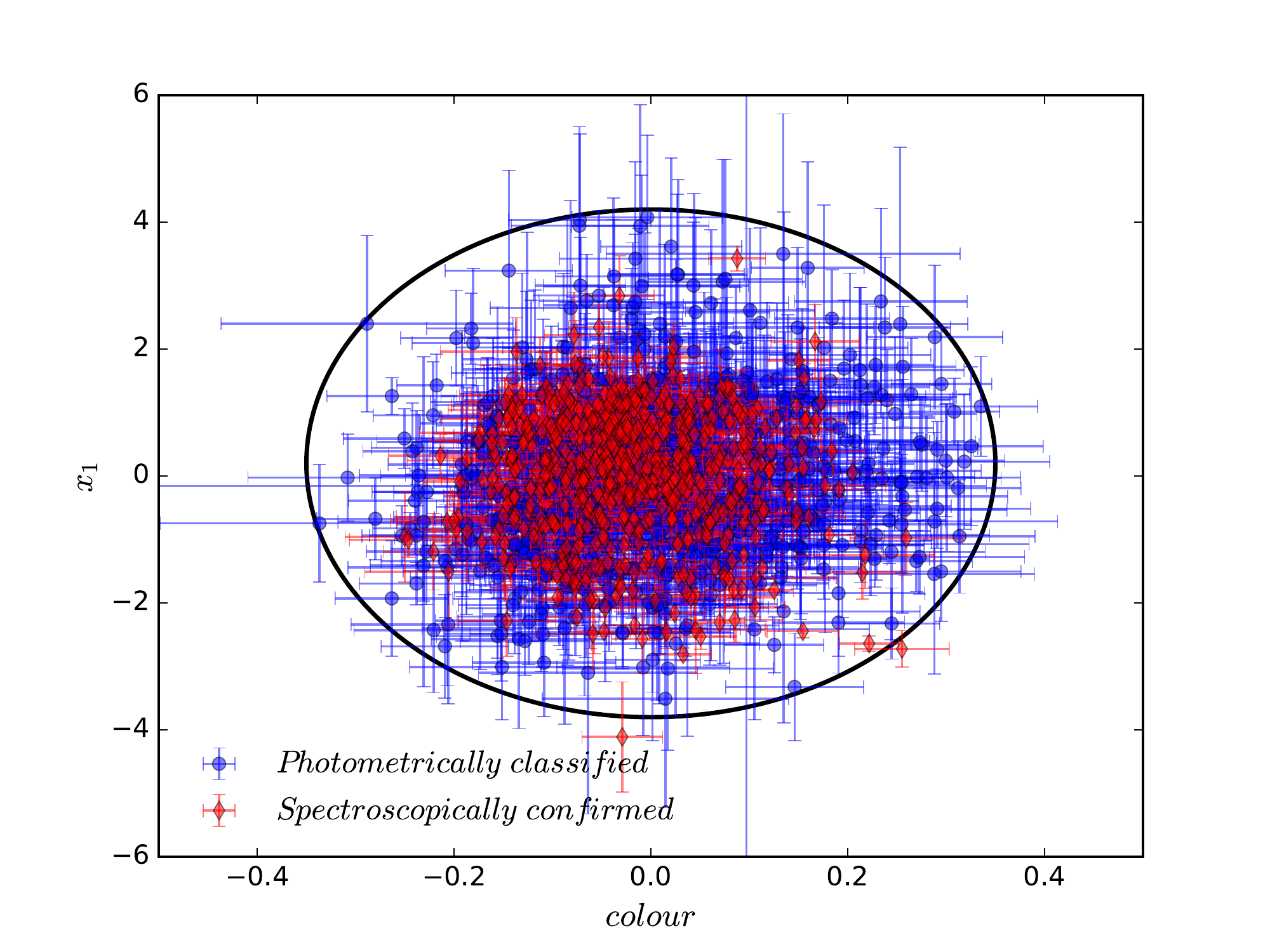}\\
\caption{Stretch-color cut. Blue circles are the photometrically classified SNe Ia that are selected after applying  the ellipsoidal cut. The boundary of the ellipse is black. For comparison we also show spectroscopically confirmed SNe Ia as red diamonds. Note how the scatter in the $x_1-c$ plane is smaller for spectroscopically confirmed SNe Ia.}
\label{cut1}   
\end{center}
\end{figure}

Two more selection steps are required for keeping the purity of objects selected as SNe Ia high. The first of these is a stretch-color ($x_1-c$) cut. This selection is motivated by inspecting the distribution of spectroscopically confirmed SNe Ia in the $x1-c$ plane. SNe Ia that are spectroscopically confirmed are tightly concentrated in the central region of the $x_1-c$ plane (see Fig.~\ref{cut1}). Nearly all of the spectroscopically confirmed SNe Ia can be enclosed by an ellipsoidal region. On the other hand, photometrically identified SNe Ia are more scattered in the $x_1-c$ plane. The scatter can be due to 1) contaminations from non-SNe Ia  and 2) larger photometric errors as SNe Ia in the photometrically classified sample are, on average, more distant and therefore fainter.

We apply an ellipsoidal cut of the form of 

\begin{equation}
\bigg(\frac{x_1}{a}\bigg)^2 + \bigg(\frac{c}{b}\bigg)^2 <1
\end{equation}

to remove outliers. Here $a$ and $b$ are the semi-major and semi-minor axes of the ellipse. They can be defined from observation or from simulation. For our sample, we have used $a=4$ and $b=0.35$, which we adopt from \cite{bazin11}. The stretch-color ellipsoidal cut is shown in Fig.~\ref{cut1}. 

\begin{figure}[htp]
\includegraphics[width=\columnwidth]{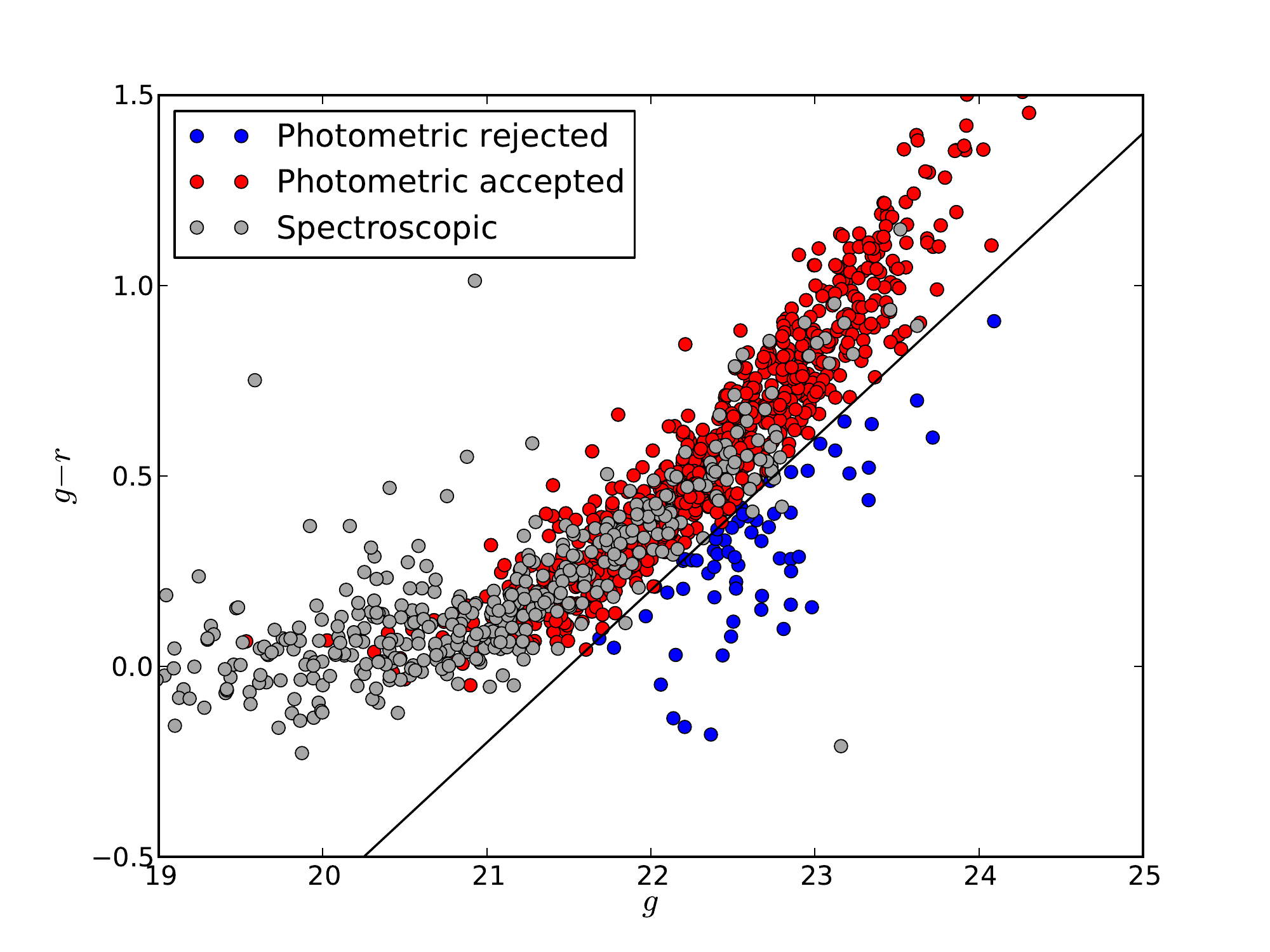}     
\caption{Rest-frame color-magnitude cut for the SDSS-II photometrically classified SN Ia sample. Blue dots are the SNe Ia that are rejected after the color-magnitude cut is applied. We draw a line according to Equation \ref{gr} that approximately separates fainter and redder objects from those that form a band in the color-magnitude plane. Red dots are the photometrically classified SNe Ia that we keep. For comparison we show spectroscopically confirmed SNe Ia from SDSS-II as grey dots.}
\label{cut2}     
\end{figure}

The second cut is applied to a color and magnitude diagram. Observations show that the SNLS SNe Ia that are spectroscopically confirmed form a thin band in color-magnitude space, such as the one shown in Fig.~\ref{cut2}. Objects below this thin band are more likely to be core-collapse SNe. For the SNLS sample, \cite{bazin11} showed that the color-magnitude cuts can improve the purity of SNe Ia by removing core-collapse SNe that were not removed from the previous ellipsoidal cut. In the SNLS sample, three sequential color-magnitude cuts are applied. These are $g-i$ vs. $g$, $r-z$ vs. $r$, and $i-z$ vs. $z$. For the SDSS-II SN Ia sample, \cite{campbell13} find that a cut in $g-r$ vs. $g$ is sufficient.

 For the SNLS photometric SN Ia sample, which comes from \cite{bazin11}, ellipsoidal and color-magnitude cuts were already applied. In this paper, we apply these two cuts only to the SDSS-II sample. An example of the color-magnitude cut is shown in Fig.~\ref{cut2}. By visual inspection we adopt the following color-magnitude criteria:

\begin{equation}
g-r < 0.4\times (g-21.5)
\label{gr}
\end{equation}

\begin{figure}[htbp]
\begin{center}
\includegraphics[width=\columnwidth]{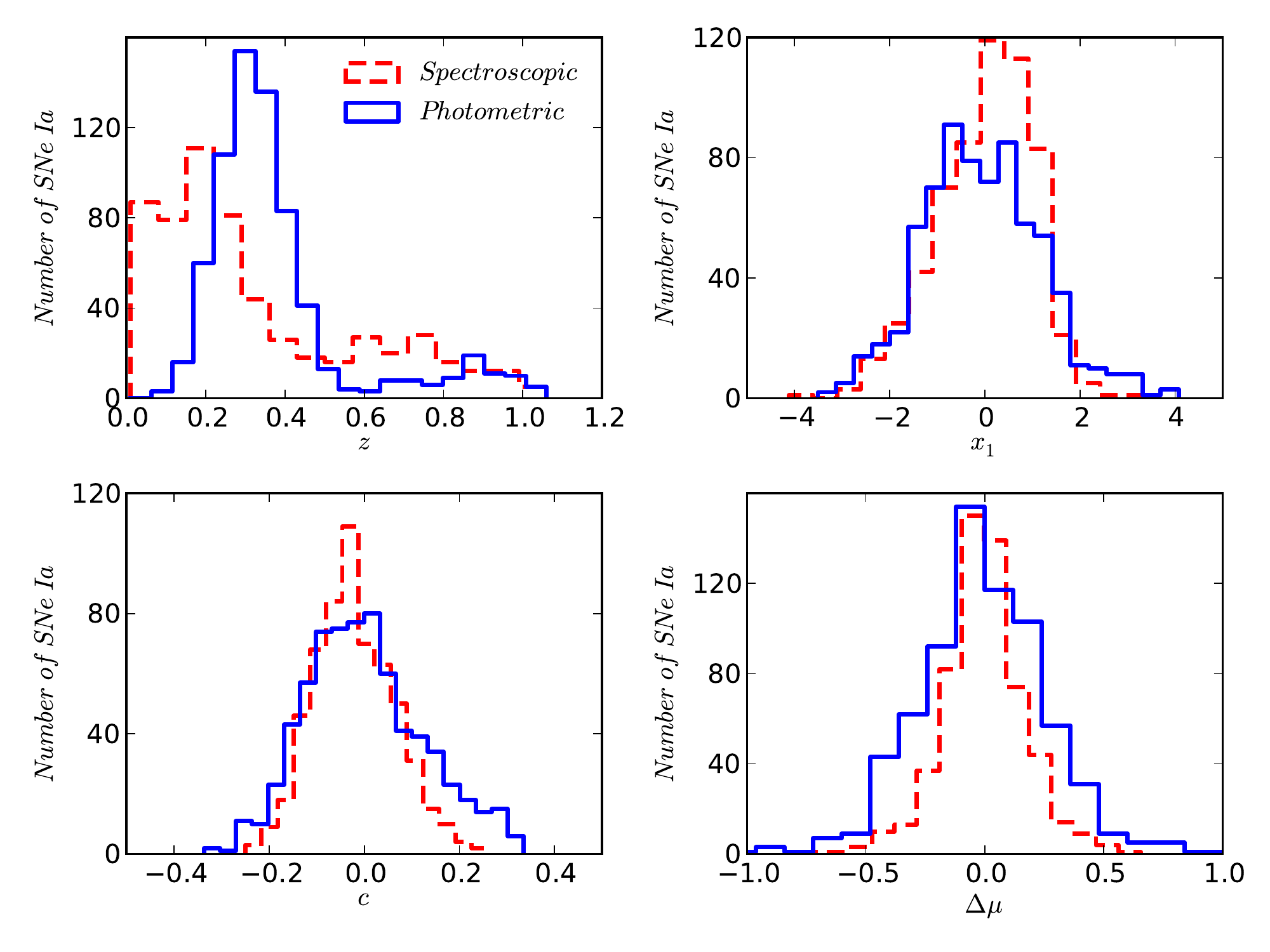}\\
\caption{Distribution of SNe Ia with redshift ($z$), light-curve width ($x_1$), color ($c$), and Hubble residual ($\Delta \mu$). Red histograms are for spectroscopically confirmed SNe Ia and blue histograms are for photometrically classified SNe Ia.} 
\label{sn}
\end{center}
\end{figure}

After these two final cuts, the number of SNe Ia from the SDSS-II photometrically classified sample drops from 944 to 661. We note that none of the photometrically classified SN Ia appears in the spectroscopically confirmed SN Ia sample. We show the distribution of SN Ia redshift, stretch, color, and Hubble residual in Fig.~\ref{sn} for both the spectroscopically confirmed and the photometrically classified SN Ia samples. 

It is clear from these distributions that the photometrically classified SN Ia sample differs from the spectroscopically confirmed SN Ia sample. SN Ia in the photometrically classified sample are on average at larger distances. This is because it is more difficult to obtain the spectra of SNe Ia at higher redshifts. 


SNe Ia in the photometrically classified sample are on average redder. Due to Malmquist bias, there is a bias against selecting redder and fainter SNe Ia in the spectroscopically confirmed sample. This bias is less for the photometrically classified sample, so the SNe Ia in the photometrically classified sample are therefore redder. We also find SNe Ia in the photometrically classified sample tend to have lower stretch. This can be explained in two ways: 1) due to Malmquist bias, there is a bias against selecting narrower and therefore fainter SNe Ia in the spectroscopically selected sample, and 2) it is easy to get redshifts for brighter, more massive host galaxies. As previous studies have found (and we will also see) more massive hosts contain lower stretch SNe Ia. Finally, we find a wider spread in the Hubble residual for the SNe Ia in the photometrically classified sample. This follows from the larger distance to these SNe Ia which makes them fainter and photometric uncertainties larger.

\begin{figure*}[htbp]
\begin{center}
\begin{tabular}{cc}
\includegraphics[width=\columnwidth]{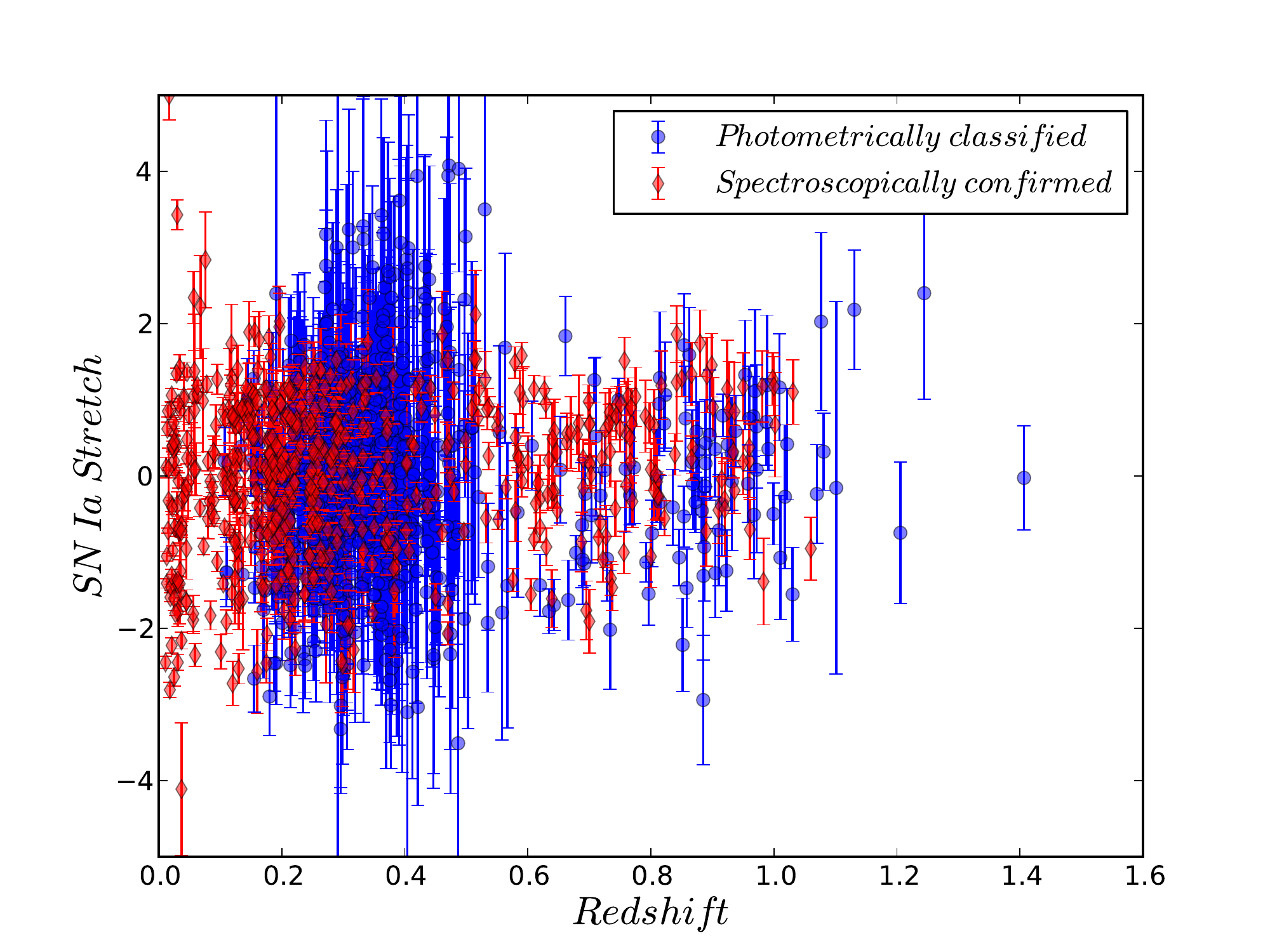} & 
\includegraphics[width=\columnwidth]{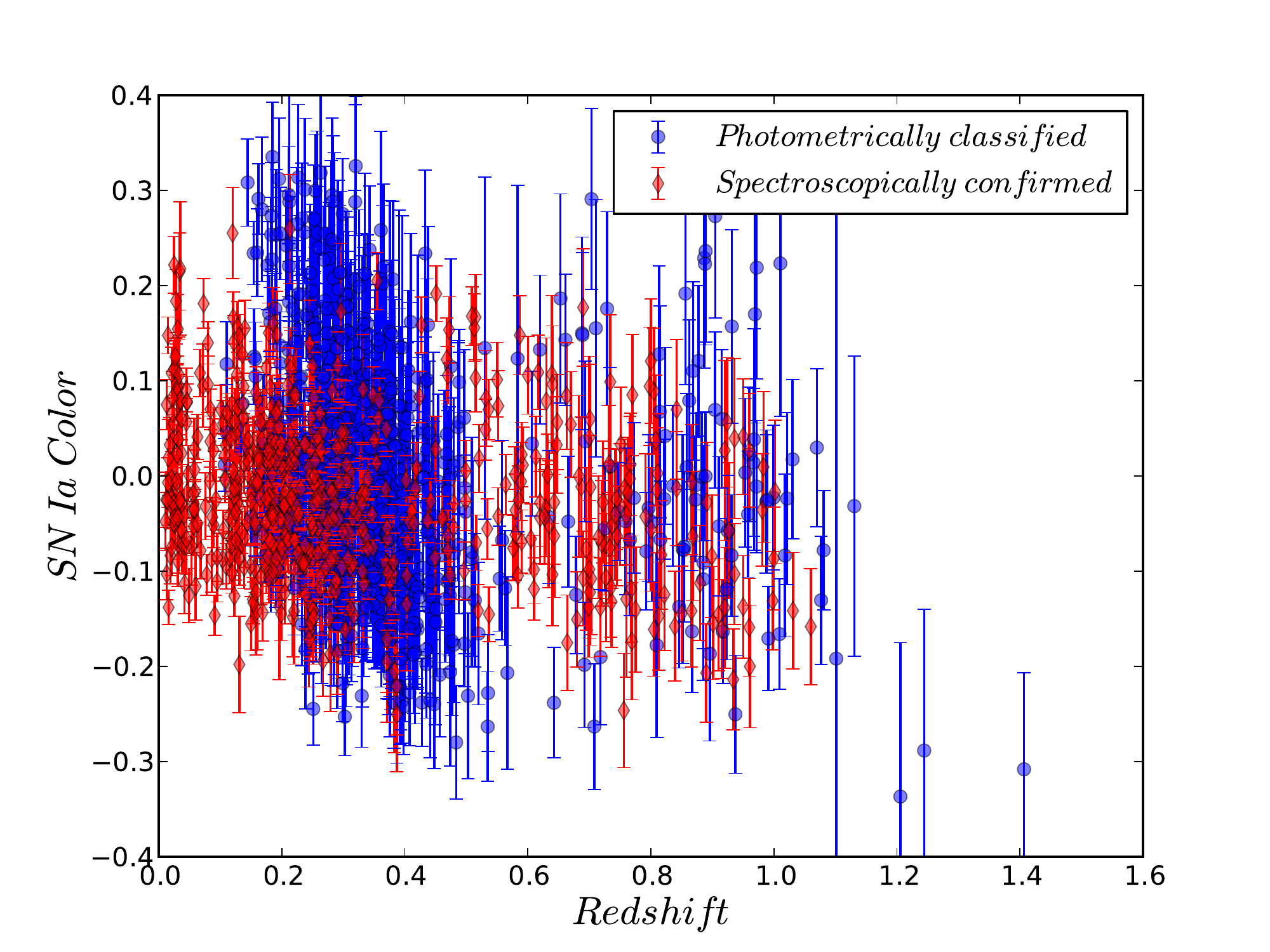}
\end{tabular}
\caption{\emph{Left}: SN Ia stretch with redshift. As we look back in time, SNe Ia tend to have larger stretch. \emph{Right}: SN Ia color with redshift. Due to Malmquist bias, we find bluer SNe Ia at larger redshifts. }
\label{vx1}
\end{center}
\end{figure*}

Finally, we show how SN Ia stretch and color varies with redshift. From Fig.~\ref{vx1} (left), we find that the SN Ia stretch tends to increase with redshift. Larger stretch means the SNe Ia brighten and fade slowly. \citet{howell07} predict a $6\%$ increase in SN Ia stretch from $z=0$ to $z=1.5$. This increase in stretch is expected from the SN Ia delay time \footnote{Delay time is the time between the formation of the SN progenitor and the occurrence of the SN.} distribution and Malmquist bias. At higher redshifts, the delay time distribution is dominated by SNe Ia that explode soon after the progenitors form (the so-called prompt SNe Ia). Prompt SNe Ia are thought to have broader light curves \citep{childress14}. We also see from Fig.~\ref{vx1} (right) that SNe Ia are bluer at higher redshifts. This is due to Malmquist bias. Bluer SNe Ia are also brighter. This trend is similar for both spectroscopically confirmed and photometrically classified SNe Ia. We also note that both stretch and color of SDSS SNe Ia that are photometrically classified ($\sim 0.2<z<0.5$) have larger scatter. This could be due to the contamination from core collapse SNe. 

We present light-curve properties in Appendix \ref{lcdata}. 

%

\subsection{Host Galaxies}

\subsubsection{Host Galaxy Identification}\label{identify}

The CfA and CSP surveys targeted SNe that were discovered in surveys that searched for transients in nearby galaxies. Therefore, host galaxies for SNe Ia that are observed in these two surveys are known a priori. But this is not the case for the SDSS-II and SNLS supernova surveys, which are rolling surveys that target `blank' regions of the sky. In these surveys, SNe Ia are discovered first and then the best matching host galaxies are identified later. SN Ia host galaxy coordinates for SDSS-II SNe Ia are available from \cite{sako14}. Host galaxy coordinates of SNLS SNe Ia are obtained from the SNLS collaboration (Hardin et al., private communication). Both SDSS-II and SNLS employ similar methods to identify host galaxies. The nearest neighbor along the line of sight is not always the correct host. 

As described in \cite{sako14}, for each SN Ia, SDSS-II begins searching for potential host galaxies within 30 arcseconds of the SN Ia using the $r$-band catalogs from DR8. The ellipsoidal shapes of these galaxies are then determined. The second moments of the $r$-band light distribution are used to derive the ellipticity and orientation of the ellipse. The semi-major axis of the ellipse is calculated from the $r$-band Petrosian half-light radius. The ellipsoidal light radius in the direction of the SN Ia is determined for each host candidate, which is termed the Directional Light Radius ($DLR$). The angular separation between SN Ia and a potential host is then normalized by this $DLR$ and this normalized distance is called $d_{DLR}$. For a given SN Ia, the $d_{DLR}$ values are ranked with increasing values and the first in the list is selected to be the host galaxy of that SN Ia.

The SNLS SN Ia host identification process is described in \cite{kronborg10} and in Hardin et al. (in preparation). It relies on two criteria. First, the normalised distances $d$ between a SN Ia and potential host galaxies are computed using the shape parameters from SExtractor \citep{bertin96}. When no galaxies are found within $d < 1.8$ of the SN Ia, the host of the SN is marked as undetected. Second, in the case of multiple host candidates, it is required that the photometric redshift of the host galaxy and the spectroscopic redshift of the SN Ia are consistent. 

\subsubsection{Host Galaxy Photometry}\label{image}
The aim of this paper is to study how the host galaxies of SNe Ia influence the properties of SNe Ia and the cosmological inferences that are made from them. Therefore, deriving host galaxy properties is an important part of this paper. Host galaxy properties are already available from previous studies. It is suitable to use them if one wants to study SNe Ia and host galaxies from a particular survey. But if one wants to combine SNe Ia and their host galaxies from different surveys, it is important that the properties of all host galaxies are derived in an consistent manner, starting from the photometry through to deriving physical properties. 

With this motivation, we analyze host galaxies for all the SNe Ia mentioned in Tables \ref{snspec} and \ref{snphoto}. In the following sections, we describe the methods that we use to obtain properties of hosts from the calibrated images.

To create a uniform sample of host galaxies, we use $ugriz$ photometry for all host galaxies in our sample. Our sample ranges from low redshift to high redshift. Low redshift host galaxies are generally bright and resolved, whereas high redshift hosts are considerably fainter and barely resolved. Therefore, for the low-redshift SN Ia hosts, single frame images have sufficient signal-to-noise for deriving photometry. For the high-redshift hosts, image stacking is necessary. 

For SDSS-II SN Ia host galaxies, we obtain calibrated image frames from DR10 of SDSS. For each SN Ia host, we obtain multiple image frames and co-add them to make a deeper image. We use the frames that have better than 2 arcseconds seeing in the $r$-band and are not contaminated by SN light. We use the IRAF\footnote{http://iraf.noao.edu/} task $imcombine$ to combine multiple images. After the co-addition is done, we align the images from all filters using the Swarp\footnote{http://terapix.iap.fr/soft/swarp/} .

For SNLS SN Ia host galaxies, deep images are constructed for each season (a season corresponds to the six consecutive months during which the field was observed). By definition, these seasonal stacks will not be contaminated by the light of SNe that occur in other seasons\footnote{This is true for most SNe. Exception include superluminous supernovae at high redshifts.}. We have obtained these stacked images from the SNLS collaboration (Hardin et al., private communication). Deep stacks are constructed by selecting 60$\%$ of the images with the best image quality. Transmission and seeing cuts (e.g. ${\rm FWHM} < 1.1$ arcseconds) are applied.  Because there are  fewer exposures in the $u$-band than in the other bands, less stringent quality cuts are applied to these images. For the Deep-D2 field, the Terapix\footnote{{http://terapix.iap.fr/}} T0006 D2-$u$ stack is used. Images are co-added using  the Swarp package to produce a  large contiguous one square degree seasonal stacks . These seasonal stacks are further co-added excluding the season during which the supernova exploded. 

For CfA3, CfA4, and CSP SN Ia hosts, we use single frame images from SDSS-II.

The resolution that we get from a ground-based astronomical image is limited by seeing. The resolution also varies from filter to filter. For ground based observations, the PSF is generally broader in the $u$-band compared to the redder bands. In order to treat images in all bands homogeneously we convolve the images so that the PSF in the convolved images matches the PSF in the worst seeing image. The images of the low-redshift hosts are not treated in this way as galaxies subtend large angles on the sky and are relatively unaffected by the changes in the PSF.
 
To accomplish the convolution process, we first create PSFs and convolution kernels for each band. We use both SExtractor and PSFEx \citep{bertin11}. Using SExtractor, we first create a catalog of point sources from an image. The configuration file and the parameter file for SExtractor are carefully set to extract only point sources. We then feed this catalog to PSFEx to produce a PSF. The next step is to create convolution kernels for the $g$, $r$, $i$, and $z$-bands by comparing the $u$-band PSF with each of the PSFs in $g$, $r$, $i$, and $z$. We use the IRAF task $psfmatch$ to create convolution kernels. 


\subsubsection{Source Extraction and  Object Catalog}\label{cat}
From host galaxy images, we extract sources and obtain their photometric and geometric properties using SExtractor. We use SExtractor in dual image mode, where a combined $ugriz$ image is used as the detection image. We include image specific quantities, such as the seeing and the zero point into the SExtractor configuration file. We also use weight images to appropriately weight regions of poor signal-to-noise. By running SExtracor on images, we obtain multi-band catalogs for host galaxies. By coordinate matching within two arcseconds, we create tables of host galaxies that include  $ugriz$ magnitudes, uncertainties, and physical parameters including half-light radius, semi-major and semi-minor axes. We use MAG$\_$AUTO, which is the magnitude determined by integrating the observed flux over an elliptical aperture that has the semi-major axis set to 2.5 times the Kron radius. 

We apply corrections for galactic extinction. Reddening estimates, $E(B-V)$, along each host galaxies line of sight are obtained from the Infrared Processing and Analysis Center$\footnote{http://irsa.ipac.caltech.edu/applications/DUST/}$ server, which uses the dust map from \cite{schlegel98}. Extinction coefficients ($R_{\lambda}$) for different bands are calculated using the York Extinction Solver$\footnote{http://www2.cadc-ccda.hia-iha.nrc-cnrc.gc.ca/community/YorkExtinctionSolver/}$ and using the Fitzpatrick extinction law \citep{fitzpatrick99}. 

%
%
%
%

\subsubsection{Host Galaxy Properties}\label{hostprop}



With $ugriz$ photometry, we derive host galaxy stellar mass and sSFR using a SED fitting code Z-PEG \citep{leborgne02}.  Z-PEG uses templates from the PEGASE2 spectral libraries \citep{fioc97} and finds the best fitting templates for host galaxies using $\chi^2$ minimization. We use the default templates in the Z-PEG code, which includes starburst, elliptical, spiral, and irregular galaxies. 

We use the \cite{rana92} IMF when deriving host galaxy properties. Host galaxy masses that we derive with this IMF match well with the recent SNLS host mass estimates (Hardin et al., in preparation). We also allow the internal extinction to vary from 0 to 0.3 in steps of 0.05 mag. The accuracy of using SED fitting techniques to obtain galaxy properties from photometry is reported for SDSS-II SN Ia hosts \citep{lampeitl10a}. On average, they find no significant difference between the galaxy masses obtained with photometry and with spectra \citep{kauffmann03}.  

\begin{figure*}[htp]
\begin{tabular}{cc}
\includegraphics[width=.5\textwidth]{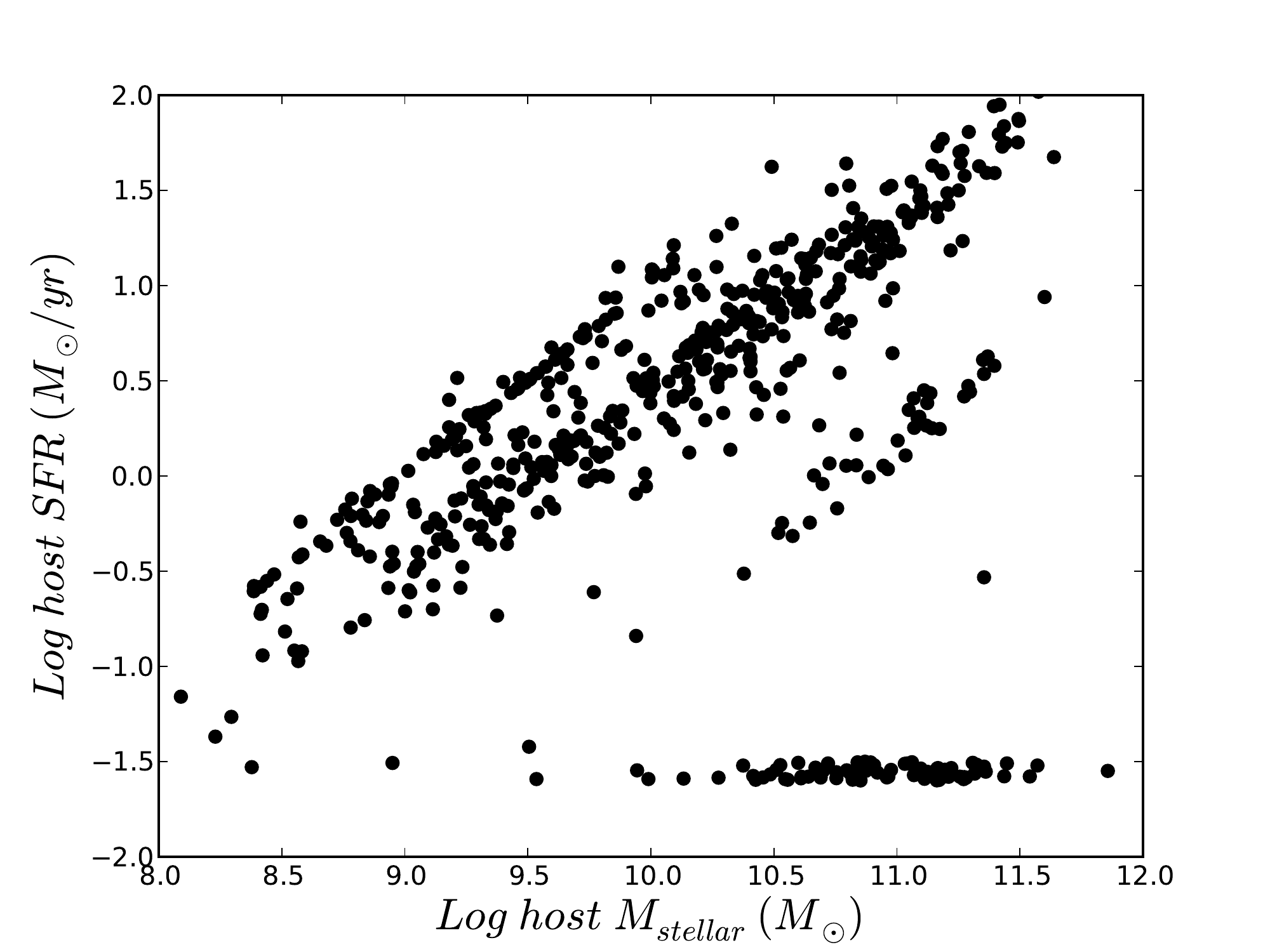}&
\includegraphics[width=.5\textwidth]{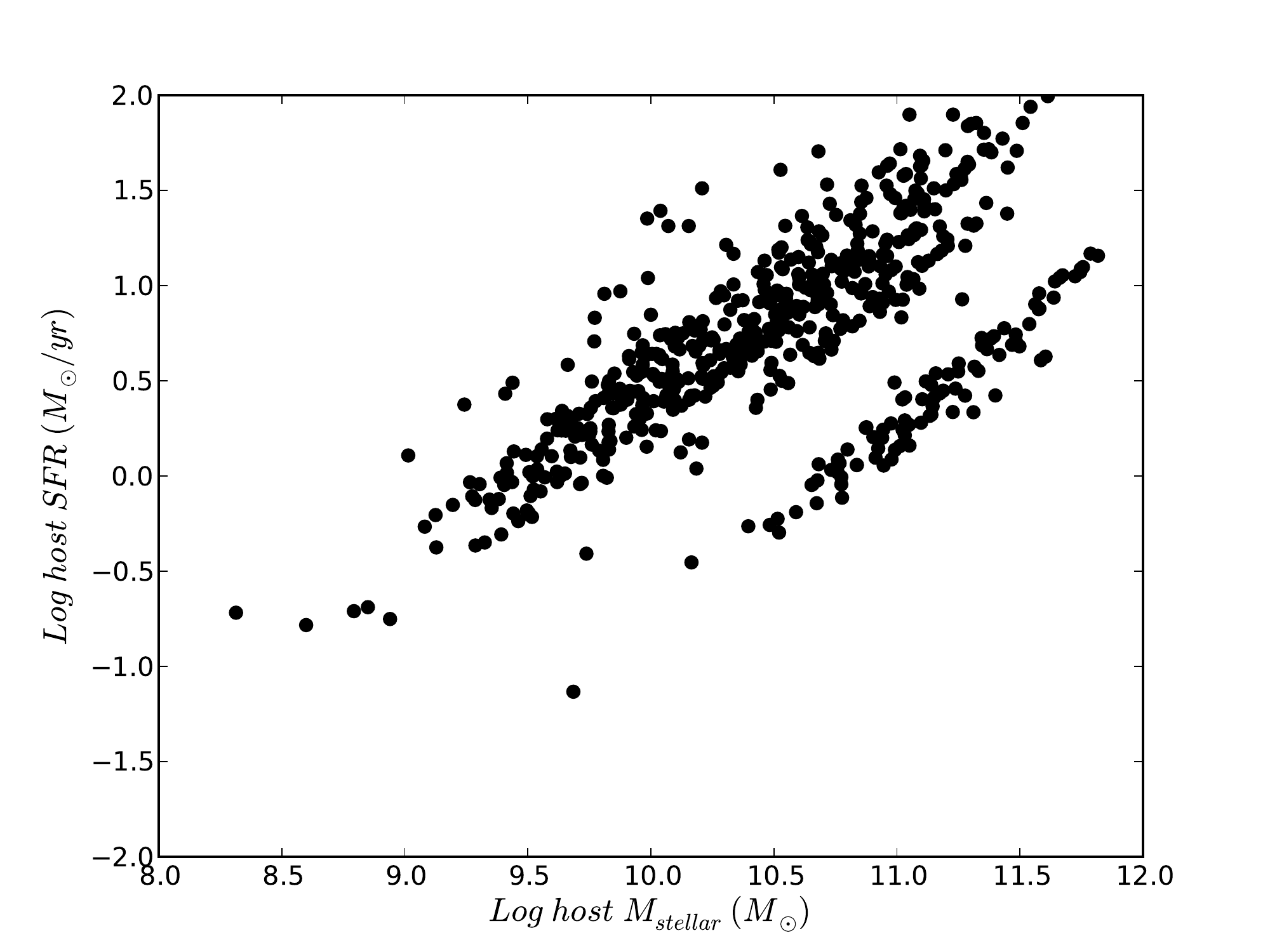}
\end{tabular}

\caption{SN Ia host galaxies in the mass-SFR plane. $Left:$ host galaxies of spectroscopically confirmed SNe Ia. $Right:$ host galaxies of photometrically classified SNe Ia. The diagonal banding in these figures comes from the limited star formation histories used in the Z-PEG templates (see text and Section \ref{lephare}). Purely passive galaxies, for which no SFR was measured, are distributed randomly between SFR of $\rm 10^{-1.7} \ to \ 10^{-1.5}  \ M_{\odot}/yr$. Note the lack of low mass hosts in the photometrically classified sample.}
\label{gmain}
\end{figure*}

In this context, we note that adding additional photometry from other bands such as ultra-violet (UV) and infra-red (IR) can improve the accuracy of inferring galaxy properties, as the SEDs become more complete. Such an improvement is shown in \cite{gupta11}. Not all of our host galaxies have UV and IR images and therefore for consistency we do not use them. We show the host galaxies in mass-SFR plane in Fig.~\ref{gmain}. Host galaxies for both the spectroscopic and photometric SN Ia samples are shown.

For some galaxies, the SFR reported by Z-PEG is zero.  They are passive galaxies with no ongoing star-formation. They are placed randomly between  $\rm Log_{10}(SFR)=-1.5$ and $\rm Log_{10}(SFR)=-1.7$ and can be seen as horizontal bands in Fig.~\ref{gmain}. 

An important feature in Fig.~\ref{gmain} is the striping event along the galaxy main sequence. Galaxies tend to line up along diagonal bands. Other studies that have used Z-PEG also found similar behavior (e.g., \citealt{sullivan10}). Using a different SED fitting code this stripping can be reduced without altering main results. We discuss this further in Section \ref{lephare}.

We note that there is a deficiency of low-mass star-forming galaxies in the photometric sample. There are two likely reasons for this. Firstly, SNe Ia from the photometrically classified sample are at higher redshifts than SNe Ia from the spectroscopically confirmed sample. Hence, we will start to loose low mass galaxies in the photometrically classified sample first because it is more difficult to get redshifts for these galaxies. Secondly, the spectroscopically confirmed sample was defined using the spectrum of the SNe Ia. Some of these may have occurred in low mass galaxies that are bright enough to obtain a mass but would have been too faint to get a redshift if they were in the photometrically classified sample.


We use $r$-band images to calculate the distance between the SN Ia and the centre of its host, and normalize this distance with the half-light radius of the host. We call this normalized projected distance (NPD). We note that NPD measurements can be biased at higher redshifts because the seeing disk is similar to the angular size of the galaxy. We also calculate the axial ratio\footnote{Axial ratio can be considered as the shape of a galaxy.} of host galaxies by measuring the ratio of semi-major ($a$) and semi-minor ($b$) axes. SExtractor output parameters A$\_$IMAGE and B$\_$IMAGE are used to estimate $a$ and $b$. The axial ratio measurement can also be biased at high redshift. Seeing will make more distant galaxies rounder than they actually are.

\begin{figure}[htbp]
\begin{center}
\includegraphics[width=\columnwidth]{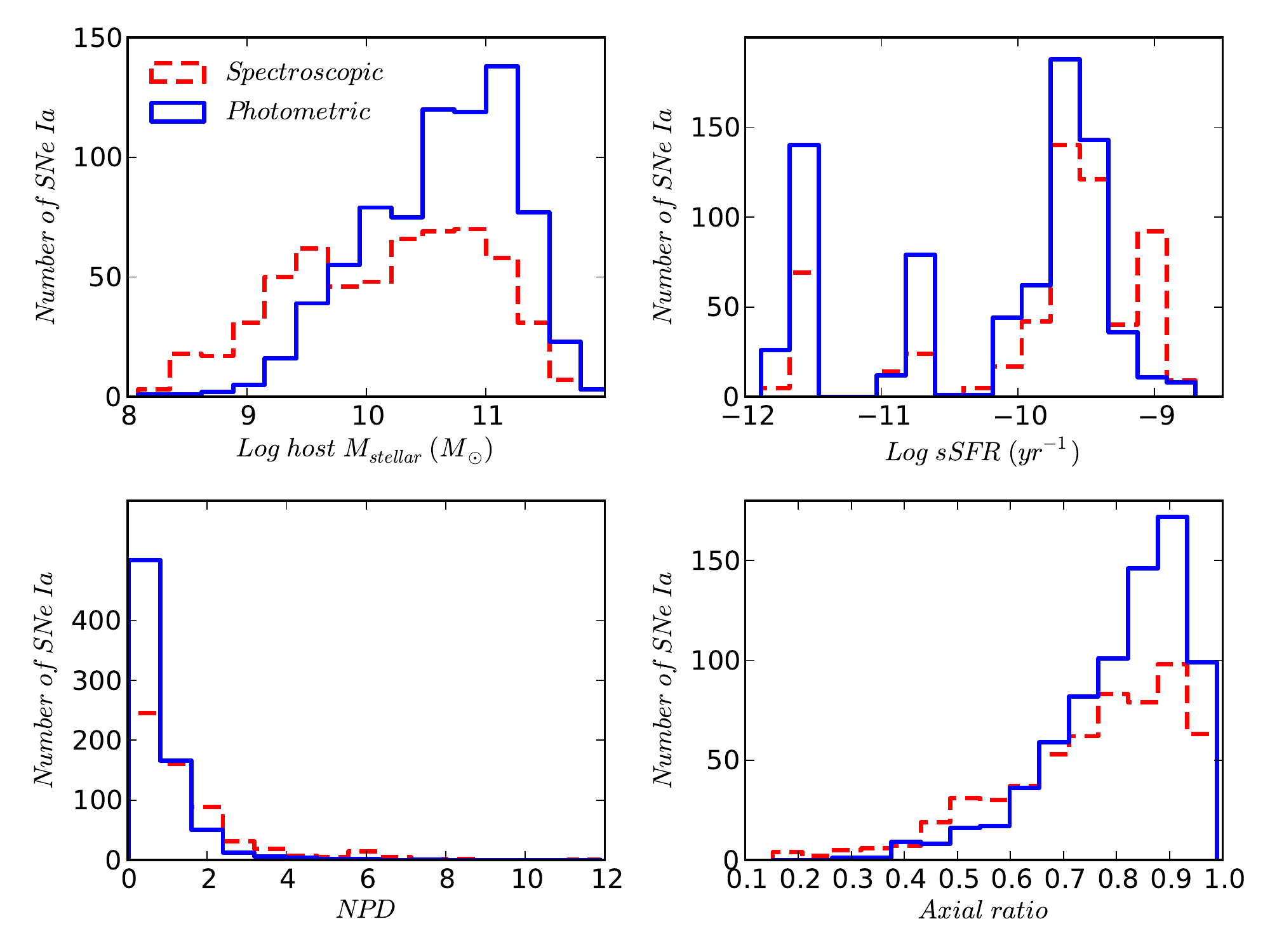}\\
\caption{Distributions of derived host properties for spectroscopically confirmed (red histograms) and photometrically classified (blue histograms) SNe Ia. We note that distributions of stellar mass are very distinct. Hosts of photometrically classified SNe Ia that have redshifts from hosts are more massive. It immediately reminds us that it is easier to get redshifts from brighter, massive hosts.}
\label{galspec}
\end{center}
\end{figure}

We use the distributions of derived host galaxy properties, as shown in Fig.~\ref{galspec}, to create SN Ia subsets that we then compare. We take the median values for host stellar mass, sSFR, and axial ratio to split the sample into paris of subsets so that an equal number of objects are in each subset. With the NPD, we split the sample at $1.00$ to separate SNe Ia inside and outside the half-light radii of host galaxies. 

As can be clearly seeing in Fig.~\ref{galspec}, the host properties of the photometrically classified sample and the spectroscopically confirmed sample differ in in terms of host stellar mass, NPD, and axial ratio. We have already noted that obtaining redshifts for the hosts of photometrically classified SNe Ia are easier when the hosts are brighter and hence massive. This explains why the hosts of the photometrically classified SNe Ia are more massive compared to the hosts of the spectroscopically confirmed SNe Ia. On average, hosts of the photometrically classified SNe Ia are rounder as they have higher axial ratios. Since these galaxies are more distant, they will appear rounder because of seeing. Seeing may also be the reason why we find the hosts in the photometrically classified sample tend to have lower NPD. 

We present physical properties of host galaxies in Appendix \ref{hostdata}. 

\section {SN Ia-Host Galaxy Correlations}\label{correlations}

After deriving SN Ia light-curves and host galaxy properties, we examine the correlations between them. We first look at the offsets in the mean properties of SNe Ia that have been split into two samples according to the properties of their hosts. Thereafter, we study how these offsets evolve with redshift.  Finally we fit cosmology to each subset independently and investigate changes in best-fit parameters. 

To investigate how SN Ia properties vary with host galaxy properties, we calculate the weighted mean of the SN Ia property that is being examined (e.g. color, or stretch) in each subset. Recall that we create SN Ia subsets based on the property of the hosts, namely host mass, sSFR and axial ratio. There is a pair of subsets for each property, split by the median value of the property. An additional pair of subsets is based on the normalized projected distance (NPD) between the SN Ia and the centre of its host. The split points are listed in Table \ref{diffcomb}.

We include intrinsic scatter of data with the individual errors when calculating the weighted mean. For every SN Ia property, we calculate the intrinsic scatter by subtracting the average error from the standard deviation of that property. For a pair of subsets, we define the offset between subsets as the difference between the weighted means. The significance of an offset is determined by adding the errors in the weighted means in quadrature. This procedure is repeated for all pairs of subsets. 

The offsets in weighted means between the pairs of subsets for $x_1$, $c$, and $\Delta \mu$ are given in Table $\ref{diffcomb}$. We plot the corresponding distributions in Figure \ref{allcorr}.

\begin{table*}[htp]
\caption{Offsets in mean SN Ia properties between the subsets in our sample. Statistical significances of offsets are shown within parentheses. The units of split points are as shown in Fig.~\ref{galspec}. }
\begin{center}
\begin{tabular}{llccc}
\hline
Property &Split Point &&Bin Difference\\
\hline
&  &$x_1$ & $c$ & $\Delta \mu$ \\
\hline
Mass & 10.53&0.44 (9.2$\sigma$)& 0.01 (0.8$\sigma$) & 0.05 (5.3$\sigma$)\\
sSFR &  -9.68&0.40 ($8.3\sigma$)&0.01 (0.5$\sigma$) & 0.03 (2.2$\sigma$)\\
NPD &1.00 &0.10 ($2.1\sigma$)  &0.01 (0.4$\sigma$) & 0.01 (0.3$\sigma$) \\
Axial  Ratio &0.82 &0.03 ($0.5\sigma$)  &0.01 (0.7$\sigma$) & 0.03 (2.5$\sigma$) \\
\hline
\end{tabular}
\end{center}
\label{diffcomb}
\end{table*}

\subsection{SN Ia Stretch and Host Properties}
In our study, we find that SN Ia stretch is correlated with host stellar mass and sSFR in such a way that low-stretch SNe Ia are hosted in high mass hosts and in hosts with low sSFRs. Between low- and high-mass galaxies, SNe Ia stretch varies by 0.44 (9.2$\sigma$) and between low- and high-sSFR it varies by 0.40 (8.3$\sigma$). Previous studies such as \cite{hamuy96}, \cite{sullivan06}, \cite{sullivan10}, and \cite{neill09} have found similar results. Stretch is independent of NPD or host axial ratio. The offsets in these cases are insignificant.

\subsection{SN Ia Color and Host Properties}
The SALT2.4 color is a combination of the intrinsic color of SN Ia and reddening from dust in the host galaxy. We see that SN Ia color does not depend on any of the host properties that we explore. \cite{childress13} report a correlation between SN Ia color and host metallicity from SuperNova Factory data, where very red ($c>0.2$) SNe Ia are found in high-metallicity hosts. The lack of clear offsets in SN Ia color between subsets is unexpected. Galaxies have metallicity gradients. The outskirts of galaxies tend to have lower metallicities. If metallicity drives SN color, the lack of an offset between samples when split by color is puzzling. Star-forming galaxies have a dusty environments, and one might expect that SNe Ia would be redder in galaxies that are dustier. More massive star-forming galaxies tend to be dustier. A recent finding from \cite{anderson15} provides partial support for this idea: they find that redder SNe Ia are associated with HII regions.

\subsection{SN Ia Hubble Residual and Host Properties}

The Hubble residual is the difference between the observed distance modulus and the best-fit model distance modulus. We define the Hubble residual, as $\Delta \mu \equiv\mu_o - \mu_m$. In this definition, a negative Hubble residual means that the SN Ia is brighter. To compute the Hubble residual, we fit our data according to the procedure that we describe later in Section \ref{ow}. Our results show that the Hubble residual varies with host mass, sSFR, and axial ratio. After correction for stretch and color, SNe Ia are brighter in high-mass galaxies by 0.05 mag (5.3$\sigma$). Our results are consistent with previous studies (e.g., \citealt{sullivan10}, \citealt{childress13}). Our results do not show a strong dependence between SN Ia brightness and the location of SN Ia with respect to its hosts. 





\begin{figure*}[htp]
\begin{tabular}{llll}
\centering
\includegraphics[width=0.33\textwidth]{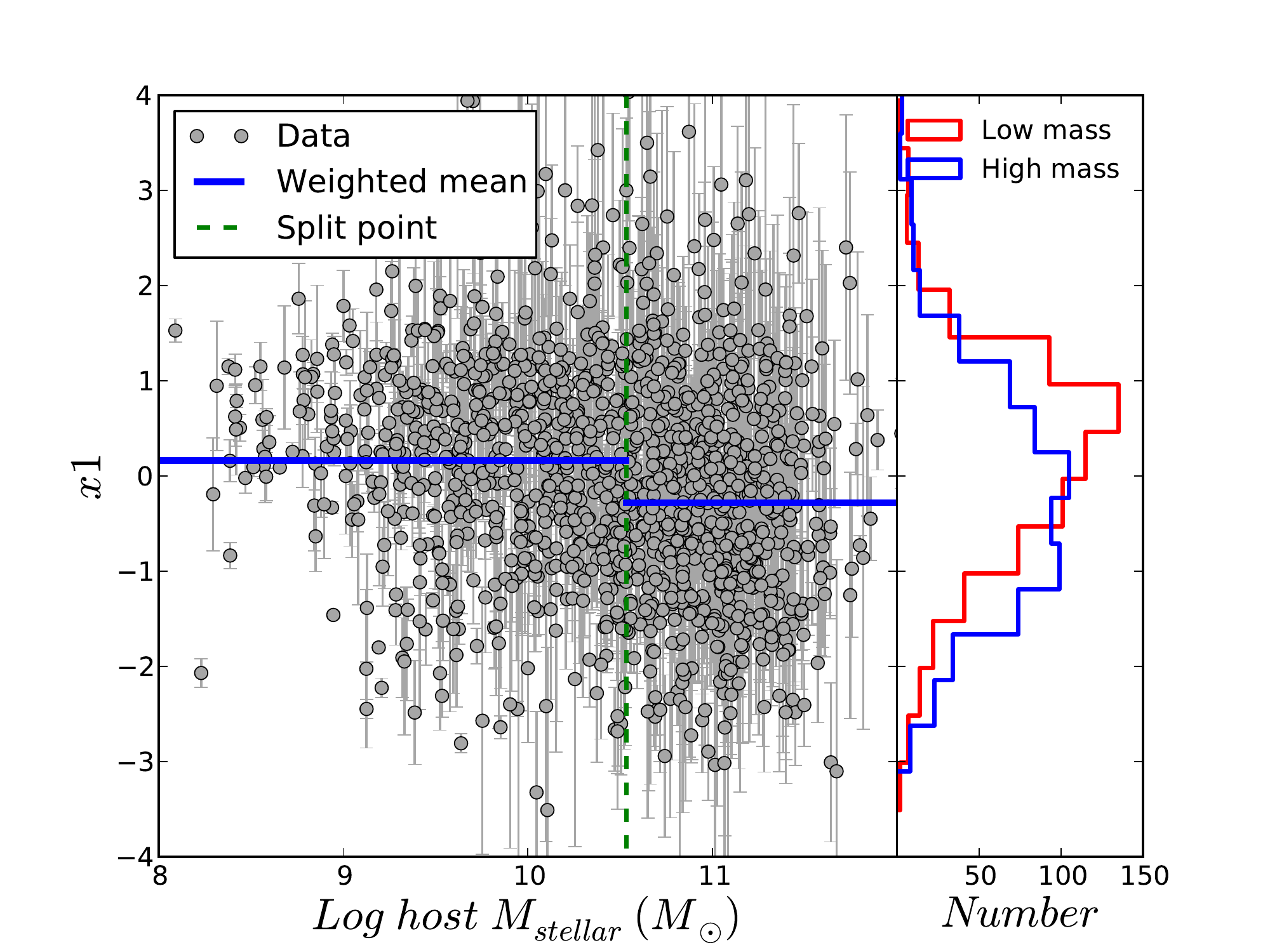}&
\includegraphics[width=0.33\textwidth]{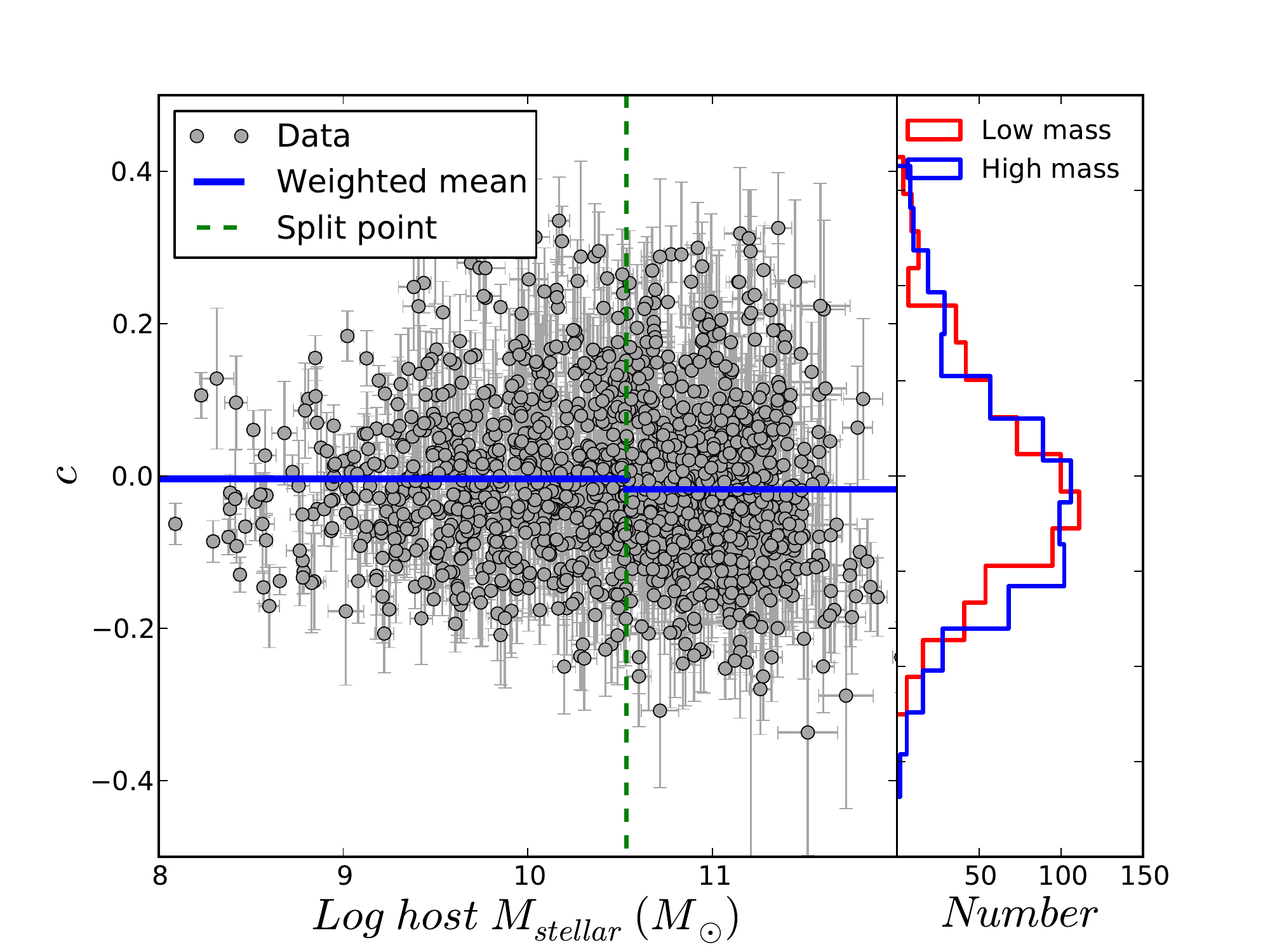}& 	
\includegraphics[width=0.33\textwidth]{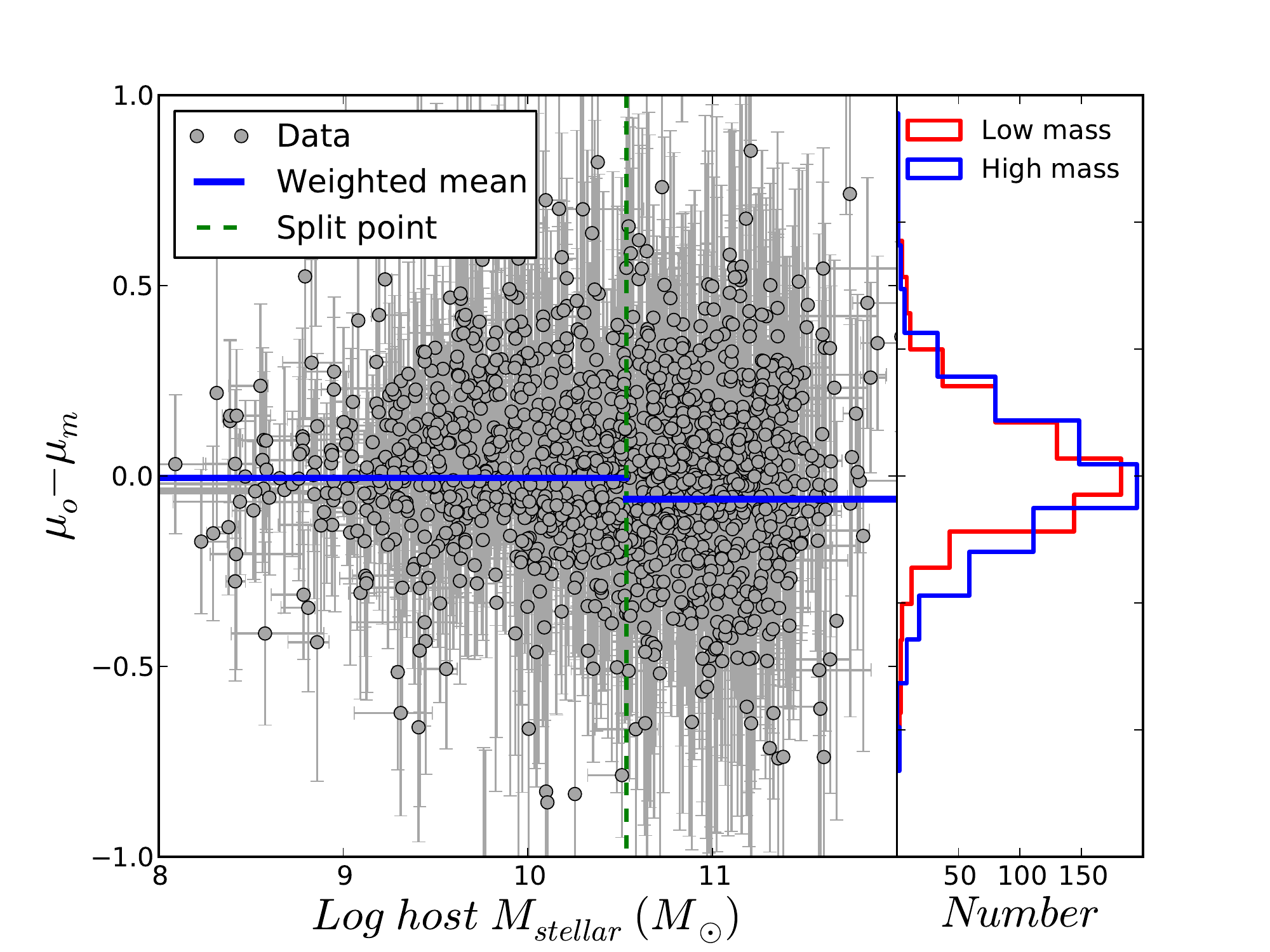}\\
\includegraphics[width=0.33\textwidth]{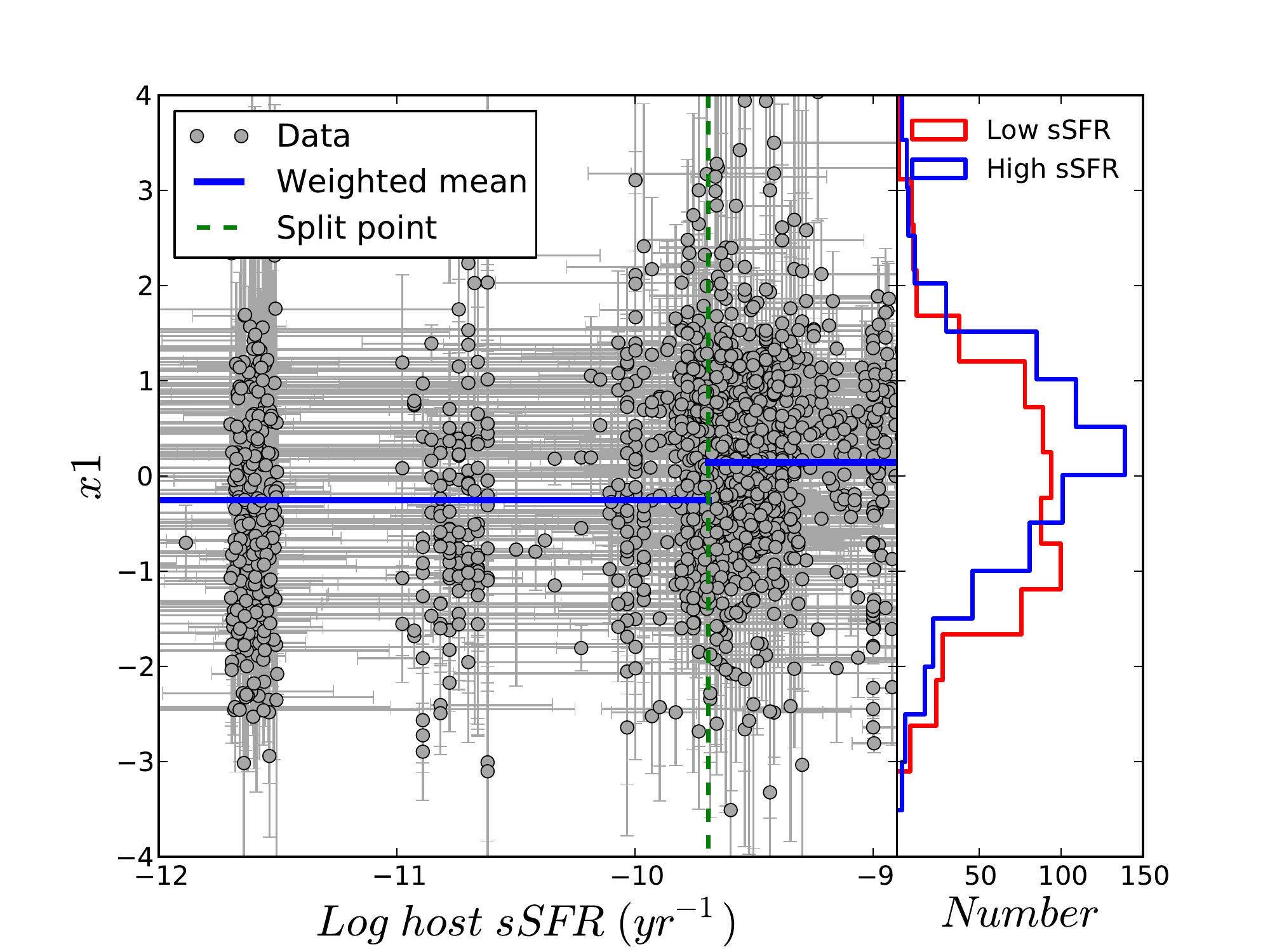}&
\includegraphics[width=0.33\textwidth]{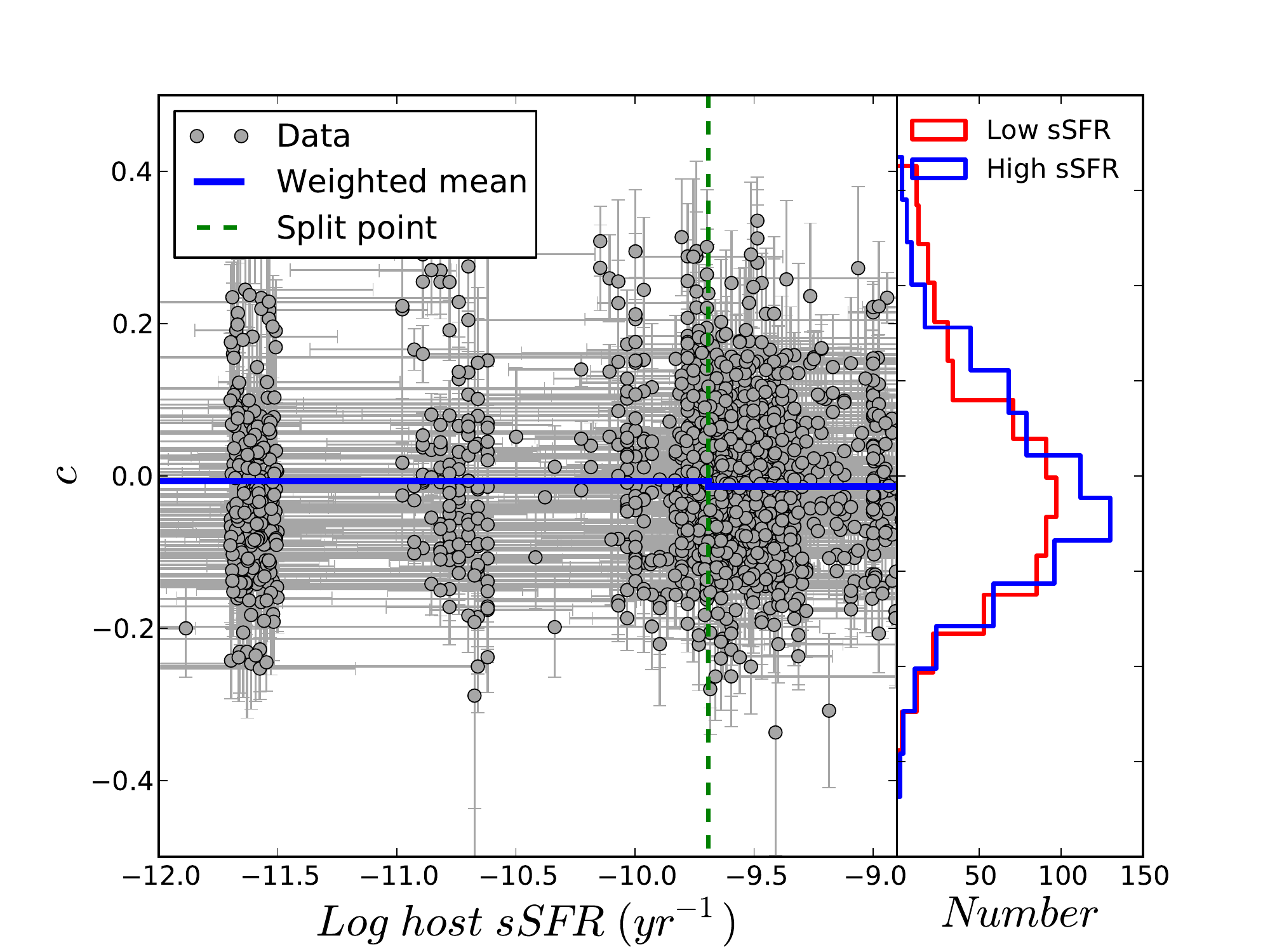}& 	
\includegraphics[width=0.33\textwidth]{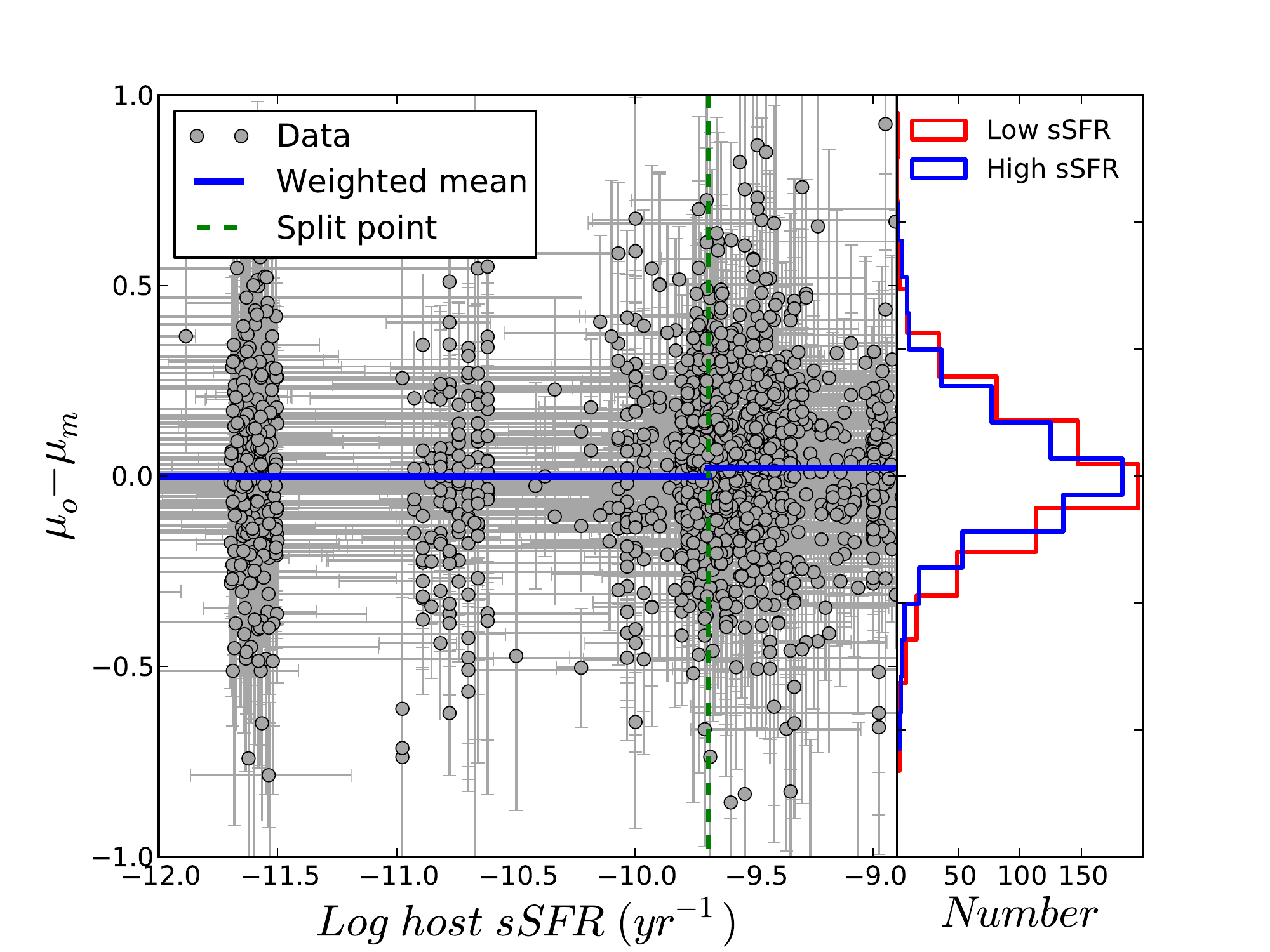}\\
\includegraphics[width=0.33\textwidth]{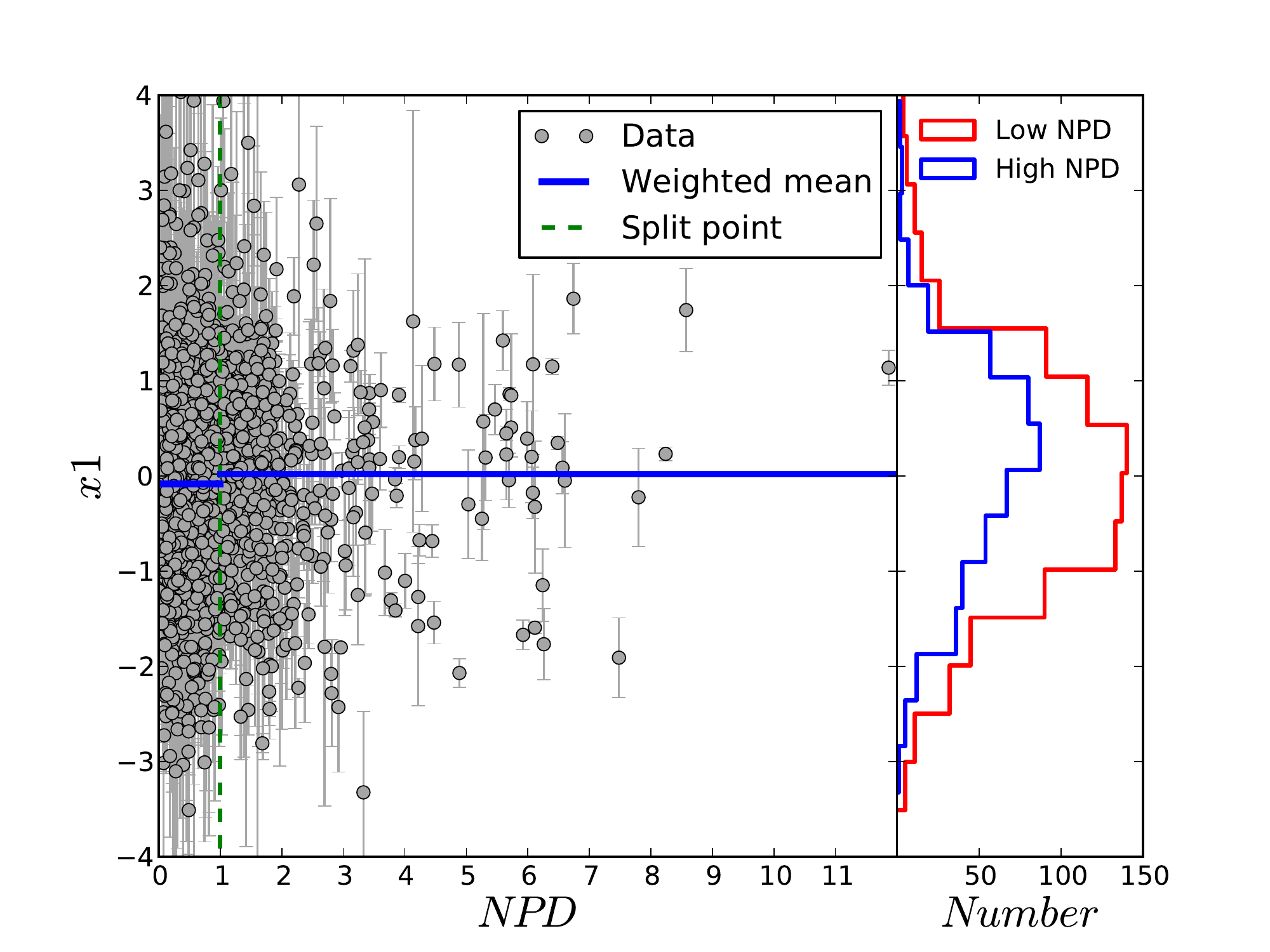}&
\includegraphics[width=0.33\textwidth]{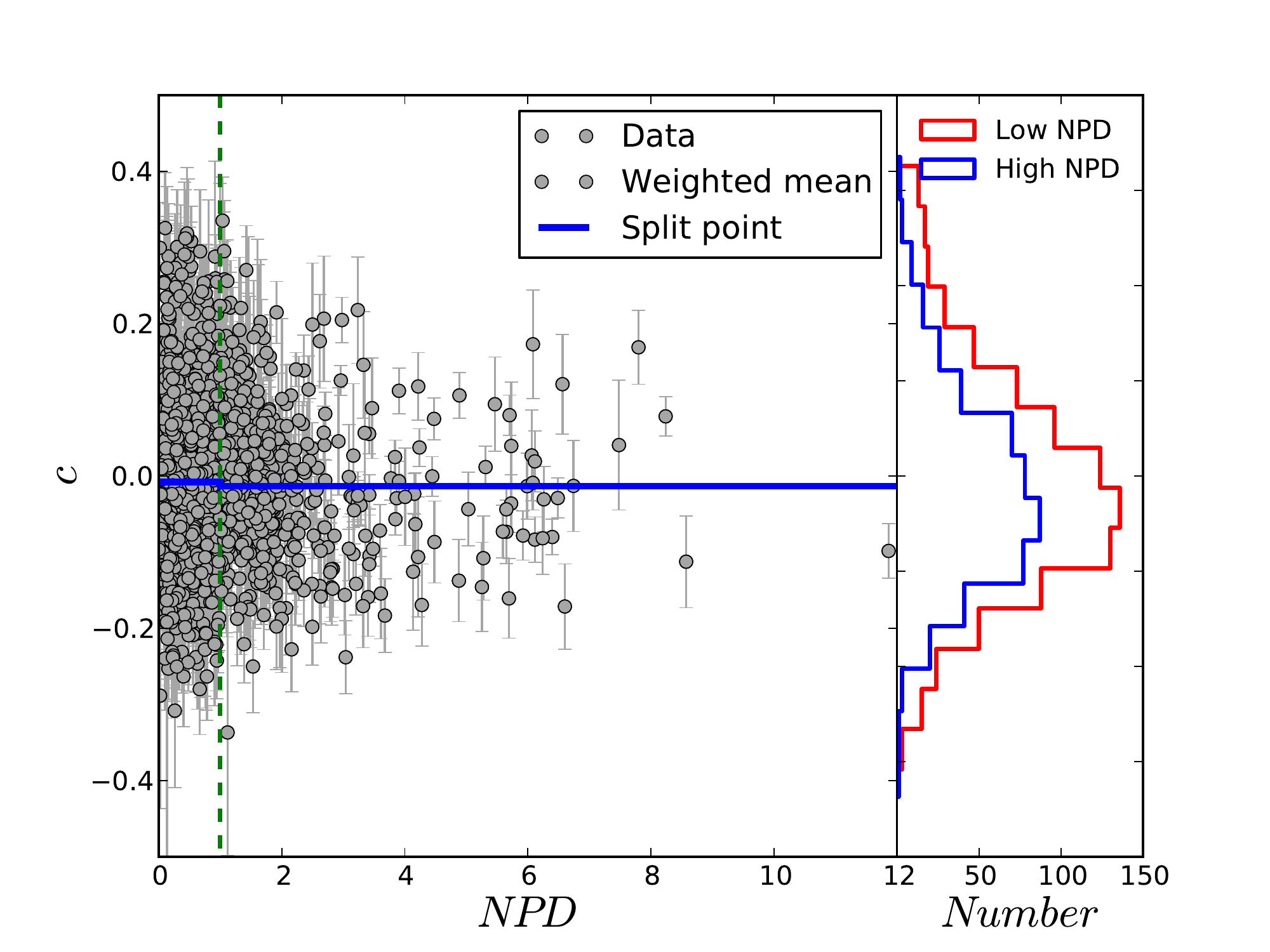}& 	
\includegraphics[width=0.33\textwidth]{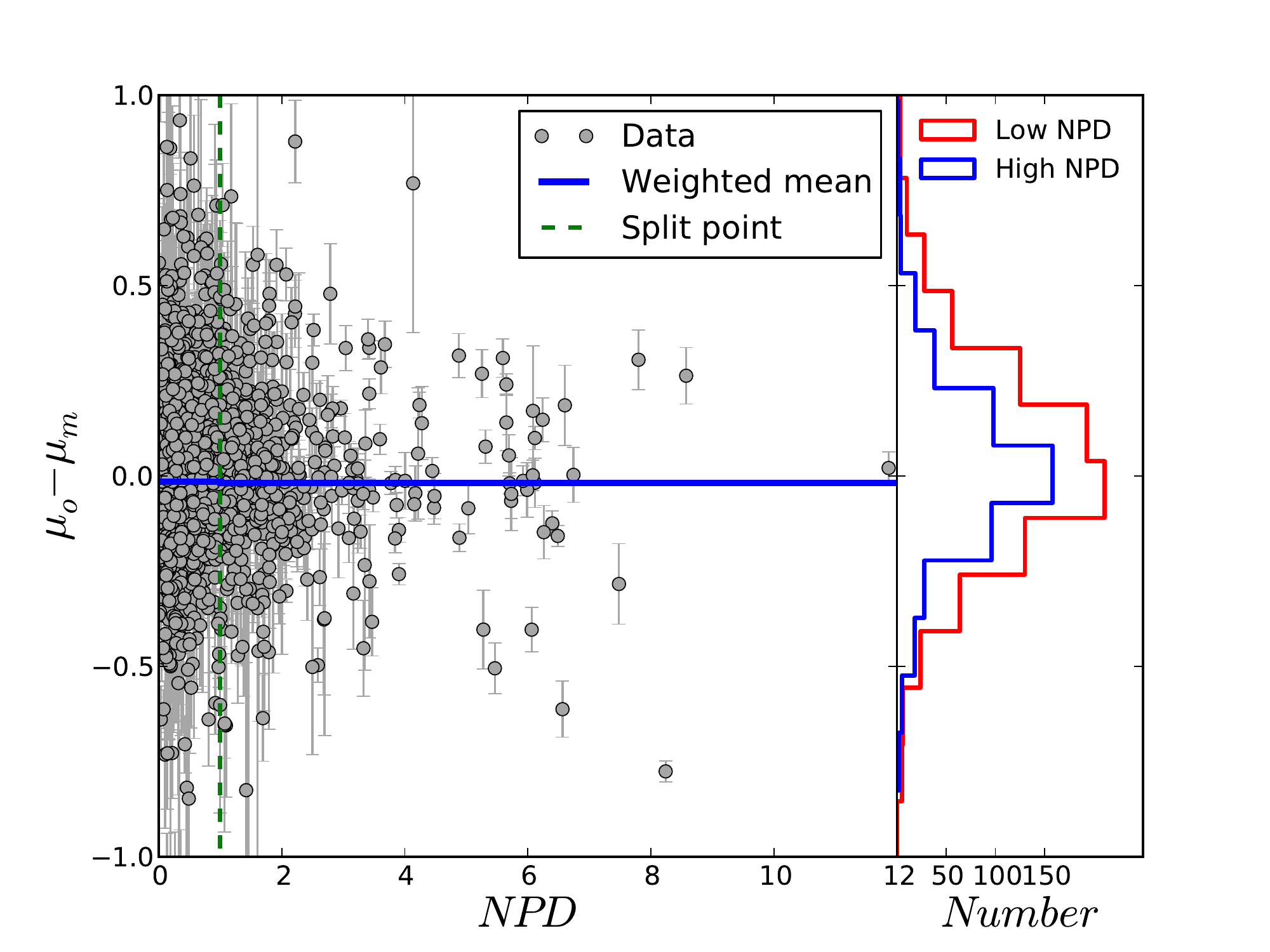}\\
\includegraphics[width=0.33\textwidth]{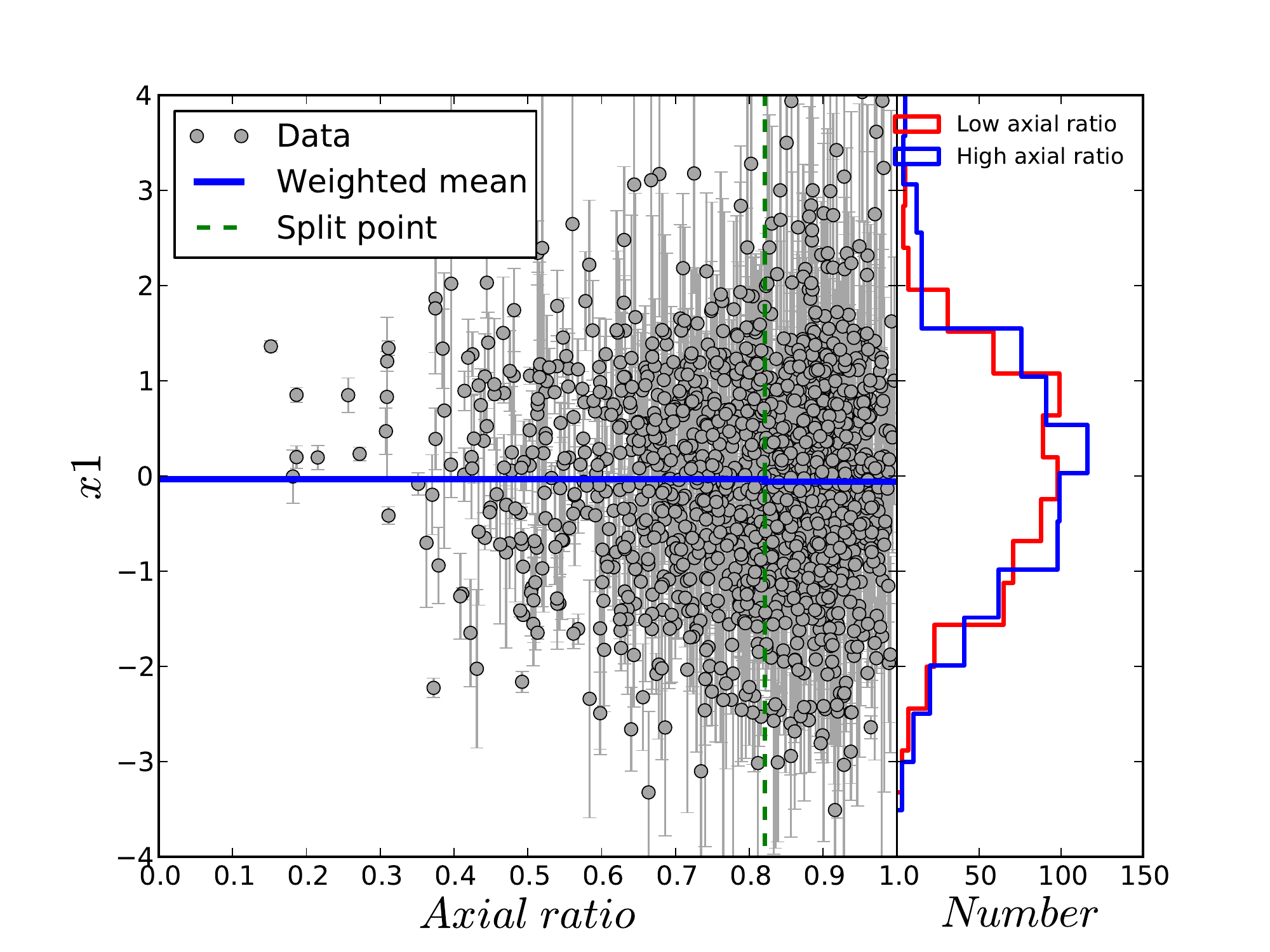}&
\includegraphics[width=0.33\textwidth]{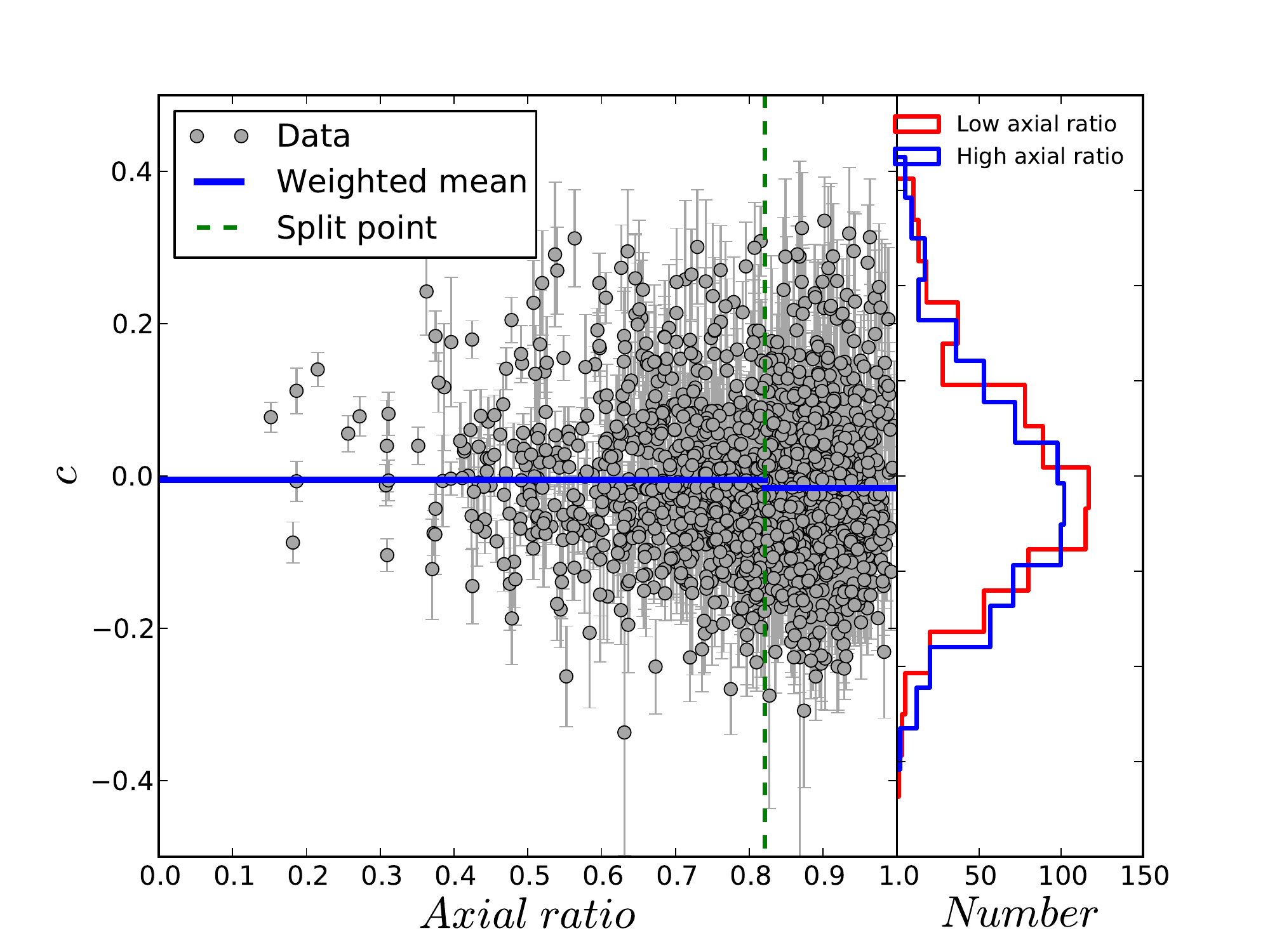}& 	
\includegraphics[width=0.33\textwidth]{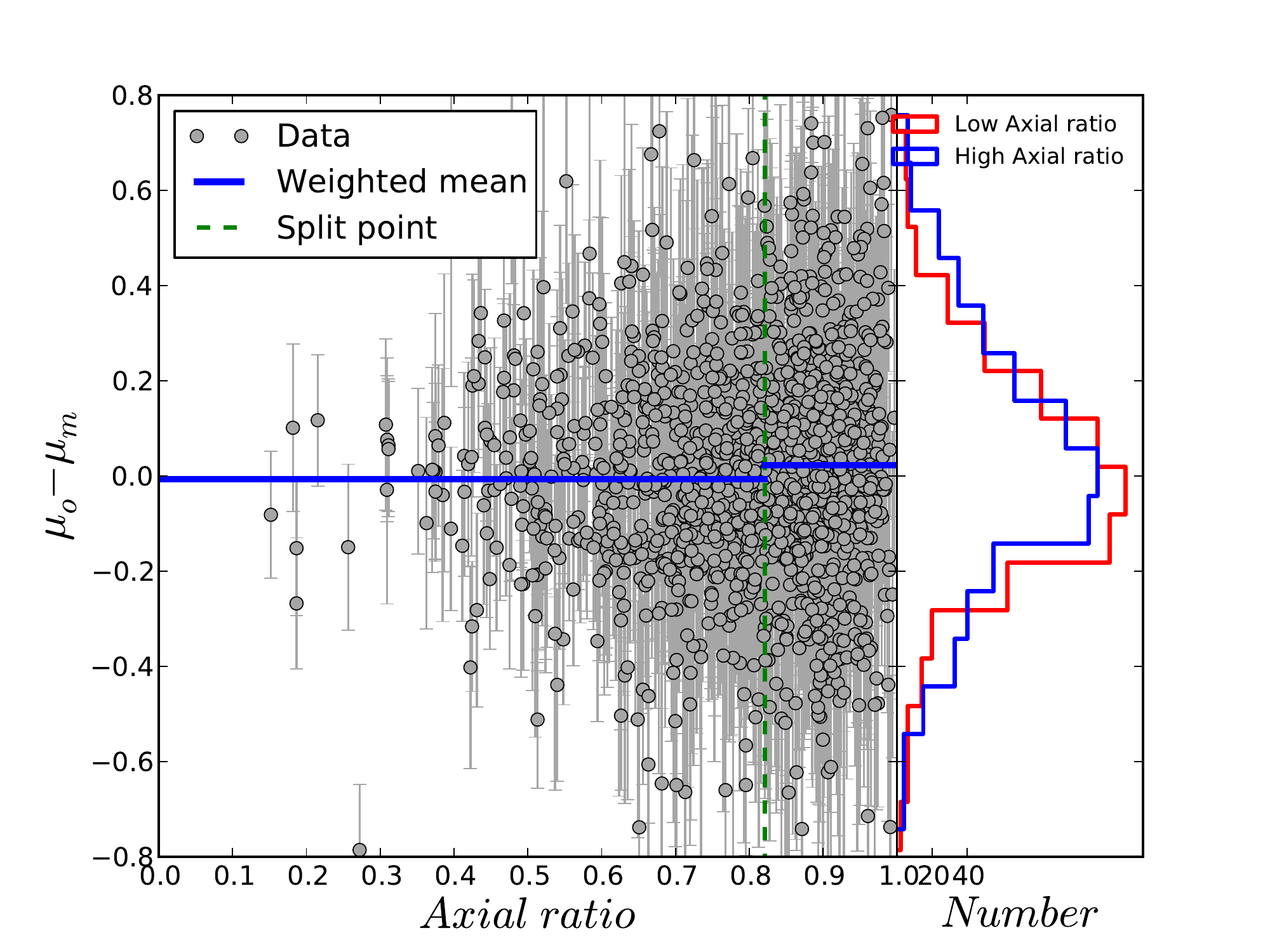}\\

\end{tabular}

 \caption{\footnotesize Dependence of SN Ia properties (left: stretch, middle: color, and right: Hubble residual) on host galaxy properties (top row: mass, second row: sSFR, third row: NPD, and bottom row: axial ratio) for our sample. On the left panel of each plot, grayscale points represent data, blue lines represent weighted means of the two subsets, and green dashed lines define the borders between subsets. On the right panel of each plot, the red and blue histograms show the stretch distribution of the two subsets. In the plots with sSFR (second row) we plot passive hosts randomly between $10^{-11.7}/yr \ and \  10^{-11.5}/yr$. In the top two left panels, it can be clearly seen that SNe Ia have significantly narrower stretches when they are hosted in galaxies that are either massive or with low rates of star formation relative to their mass (sSFR).  SN Ia stretch does not seem to depend on where SNe Ia are located with respect to  the centre of their hosts (NPD), or with the shape of their hosts (axial ratio). We find that SN Ia color does not seem to vary with any host property. On average, SNe Ia are more luminous in massive galaxies and galaxies with low sSFR. Table \ref{diffcomb} shows the significances of these results.}
\label{allcorr}     
\end{figure*}

\subsection{Distribution Comparison}\label{distri}
After studying SN Ia properties between pairs of subsets in previous sections (also see Fig.~\ref{allcorr}), we can ask the question whether the distributions of SN Ia properties and Hubble residuals in the various subsets differ. A popular test for this is the two-sided Kolmogorov-Smirnov (K-S) test. The null hypothesis is that the two groups are drawn from the same parent distribution. The $\emph{p}$-values that we calculate from the K-S test for the cumulative distributions of SN Ia color, stretch and the Hubble residual are shown in Table \ref{ks}. We can reject the null hypothesis and state that SN Ia stretch and Hubble residuals are different when SNe Ia are split according to either mass or sSFR, since $p$-values are less than $10\%$. There is also evidence that SNe Ia differ in color in the subsets that are split according to the axial ratio of the host galaxy. We find no evidence for a relationship between the properties of SNe Ia and their location with respect to the centers of their host galaxies. The impact these differences have on cosmological inference is a question that we address in Section \ref{ow}.
\begin{table}[htp]
\centering
\caption{ $D$-statistics (and $p$-values) of the K-S test.}
\begin{tabular}{lccc}
\hline
Property & $ x_1$ & $c$ & $\Delta \mu$ \\
\hline
Mass & 0.22 (0$\%$)& 0.09 (22$\%$) & 0.13 (1$\%$)\\
sSFR & 0.20 (0$\%$) & 0.03 (99$\%$) & 0.12 (1$\%$ )\\
NPD & 0.05 (77$\%$) & 0.06 (63$\%$) & 0.07 (46$\%$)\\
Axial Ratio & 0.06 (55$\%$) & 0.10 (8$\%$) & 0.08 (30$\%$)\\
\hline
\label{ks}
\end{tabular}
\end{table}%

%
%

\subsection{Redshift Evolution of the Offsets in SNe Ia Properties}\label{evolution}
We have built a SN Ia sample that have redshifts that range between $0.01<z<1.2$. This means that we are studying SN Ia properties up to the time when the universe was almost half of its present age. This allows us to study how the offsets between SN Ia subsets change with time. Since galaxies evolve with time, we can hypothesize that the offsets that we have already seen may change with redshift. 

\begin{figure*}[htp]

  \centering

  \begin{tabular}{cccc}


    \includegraphics[width=.33\textwidth]{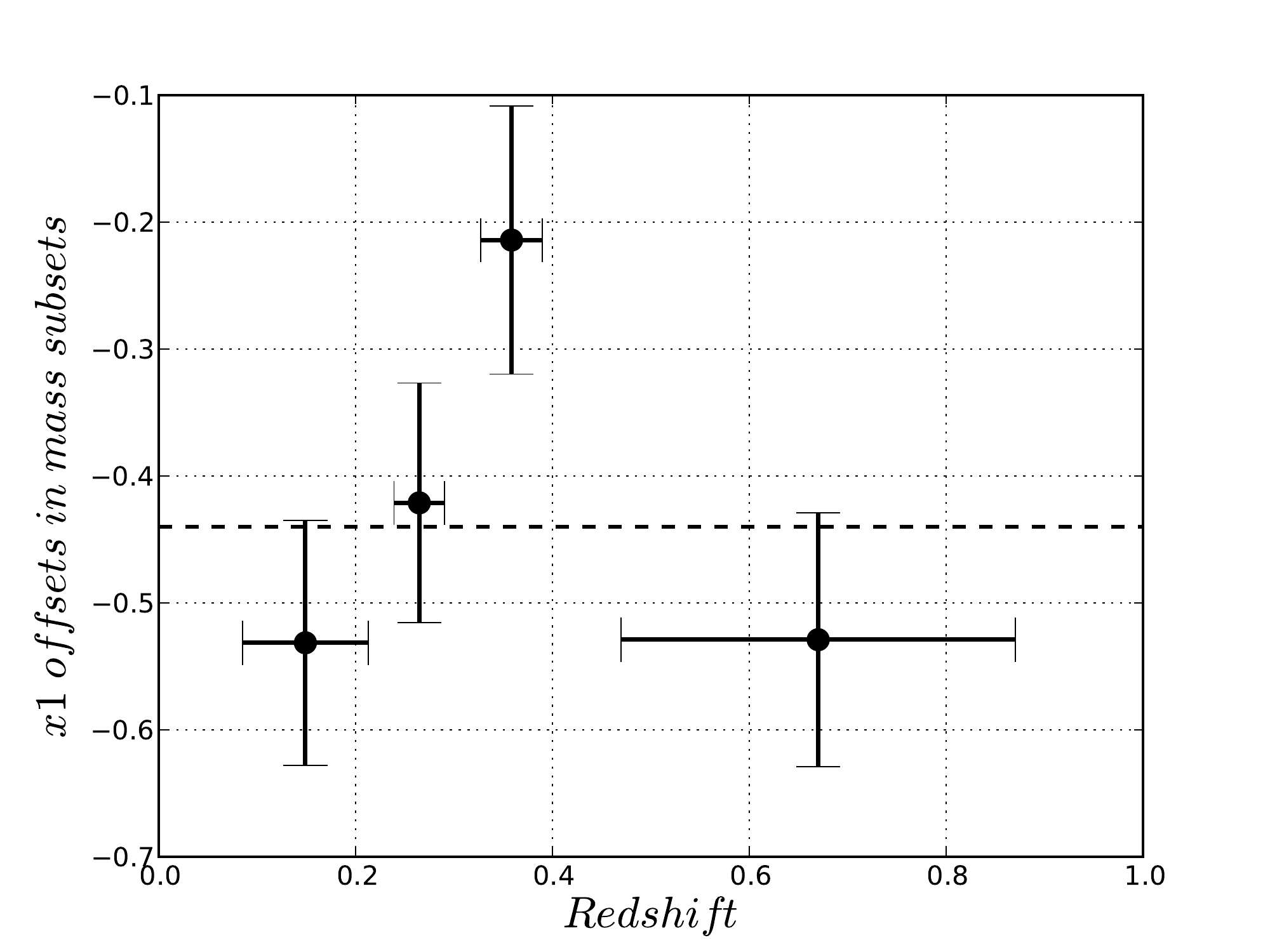}&
    \includegraphics[width=.33\textwidth]{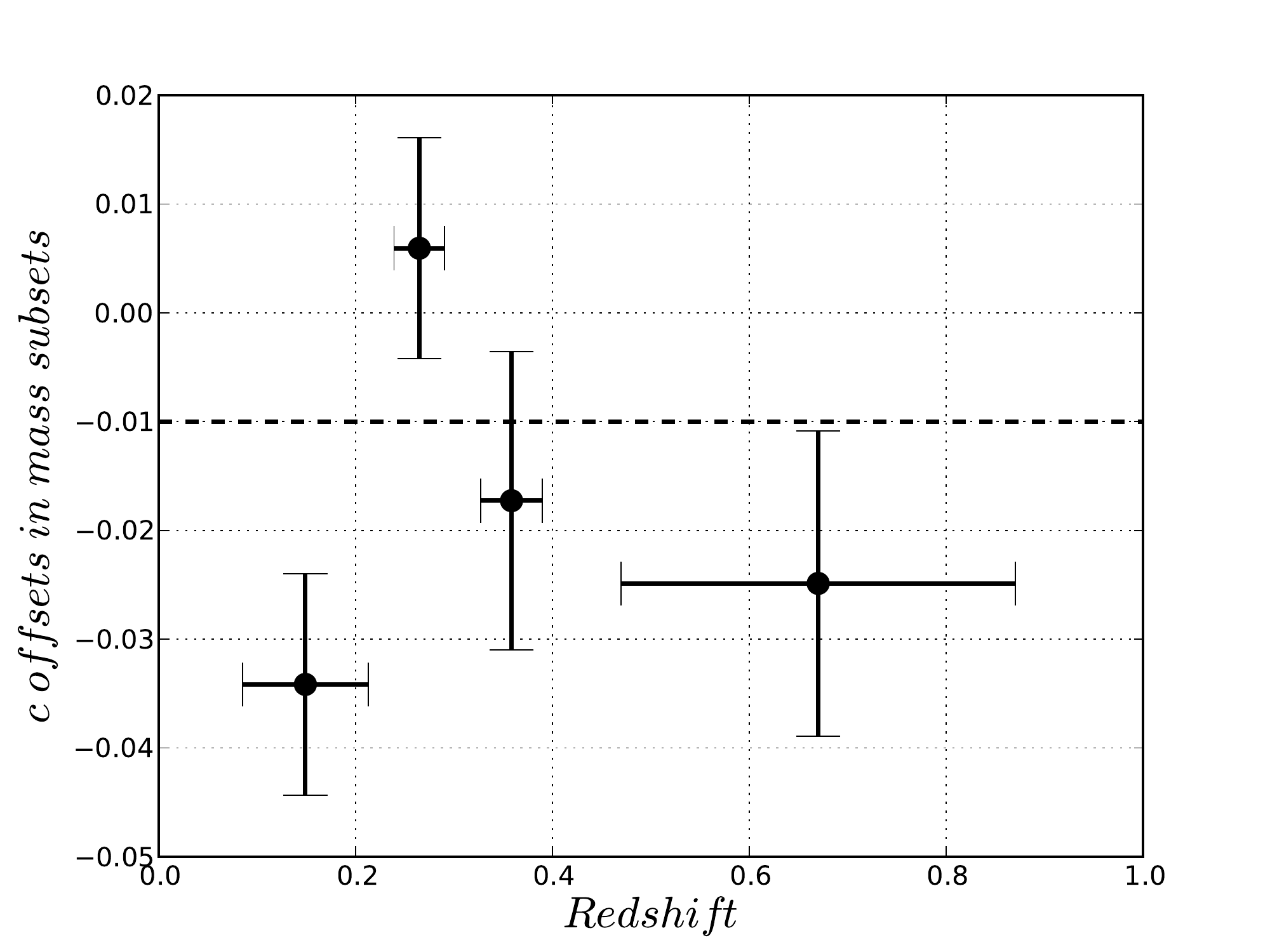}&	
     \includegraphics[width=.33\textwidth]{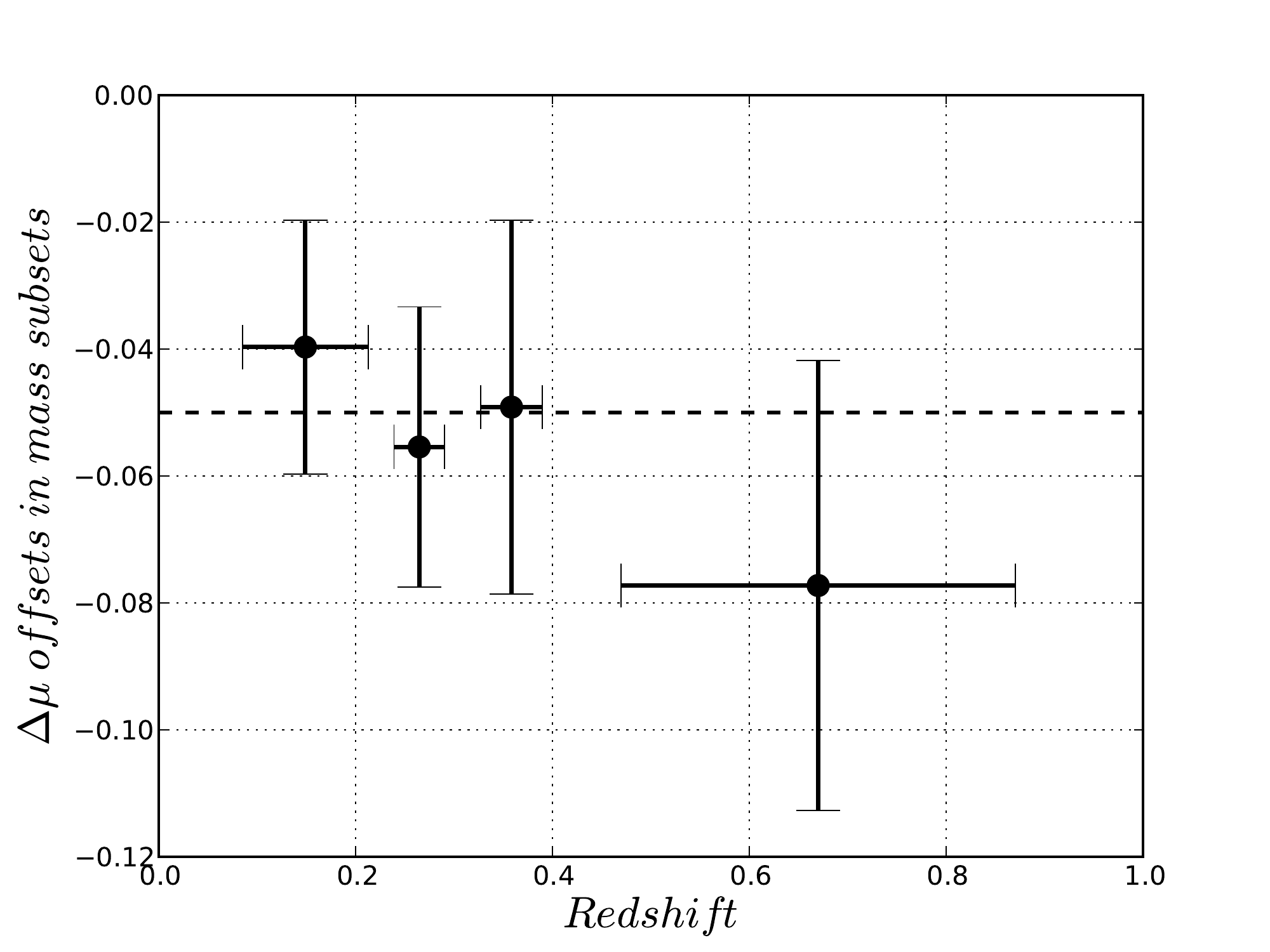}\\
      \includegraphics[width=.33\textwidth]{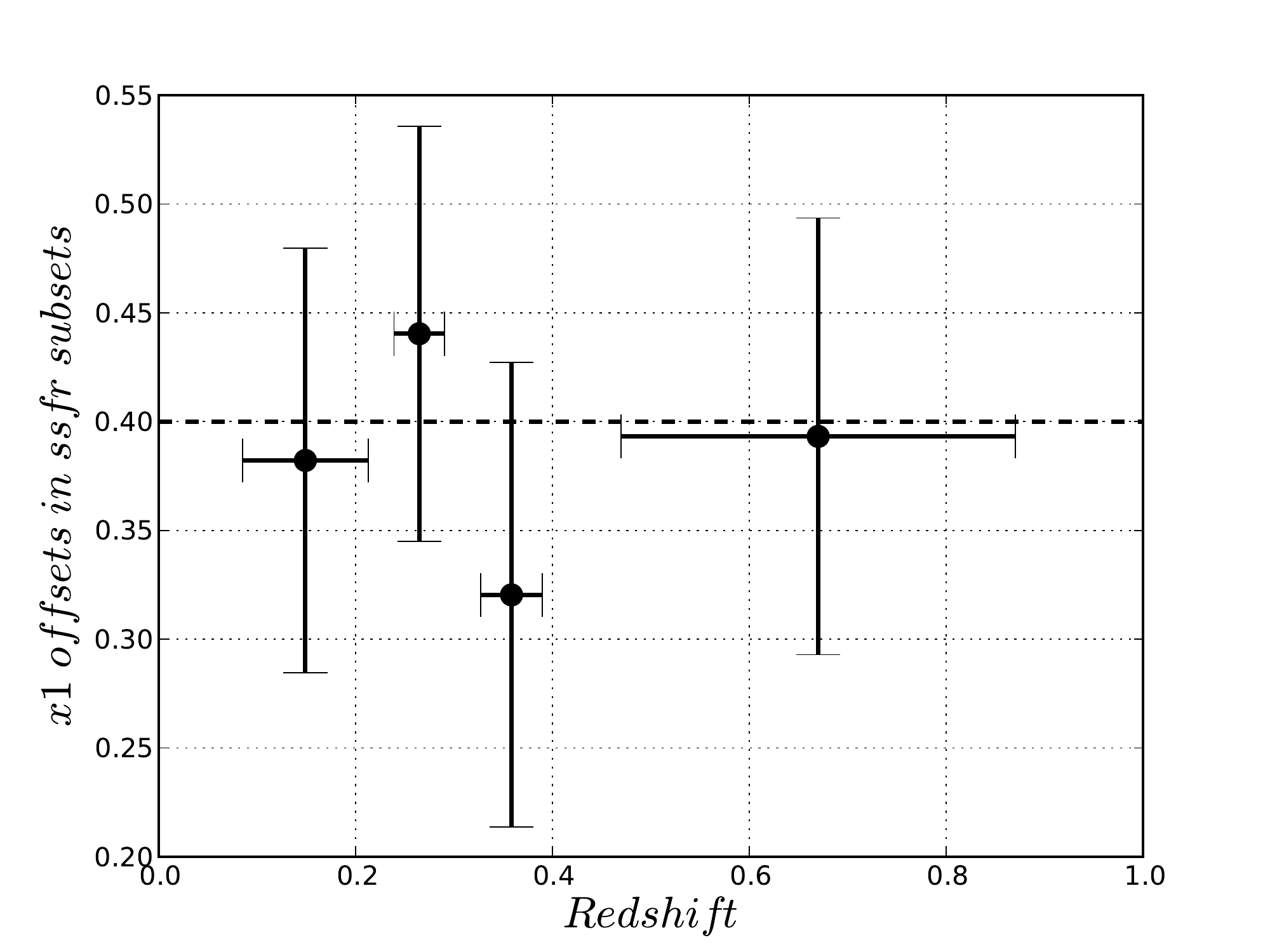}&
    \includegraphics[width=.33\textwidth]{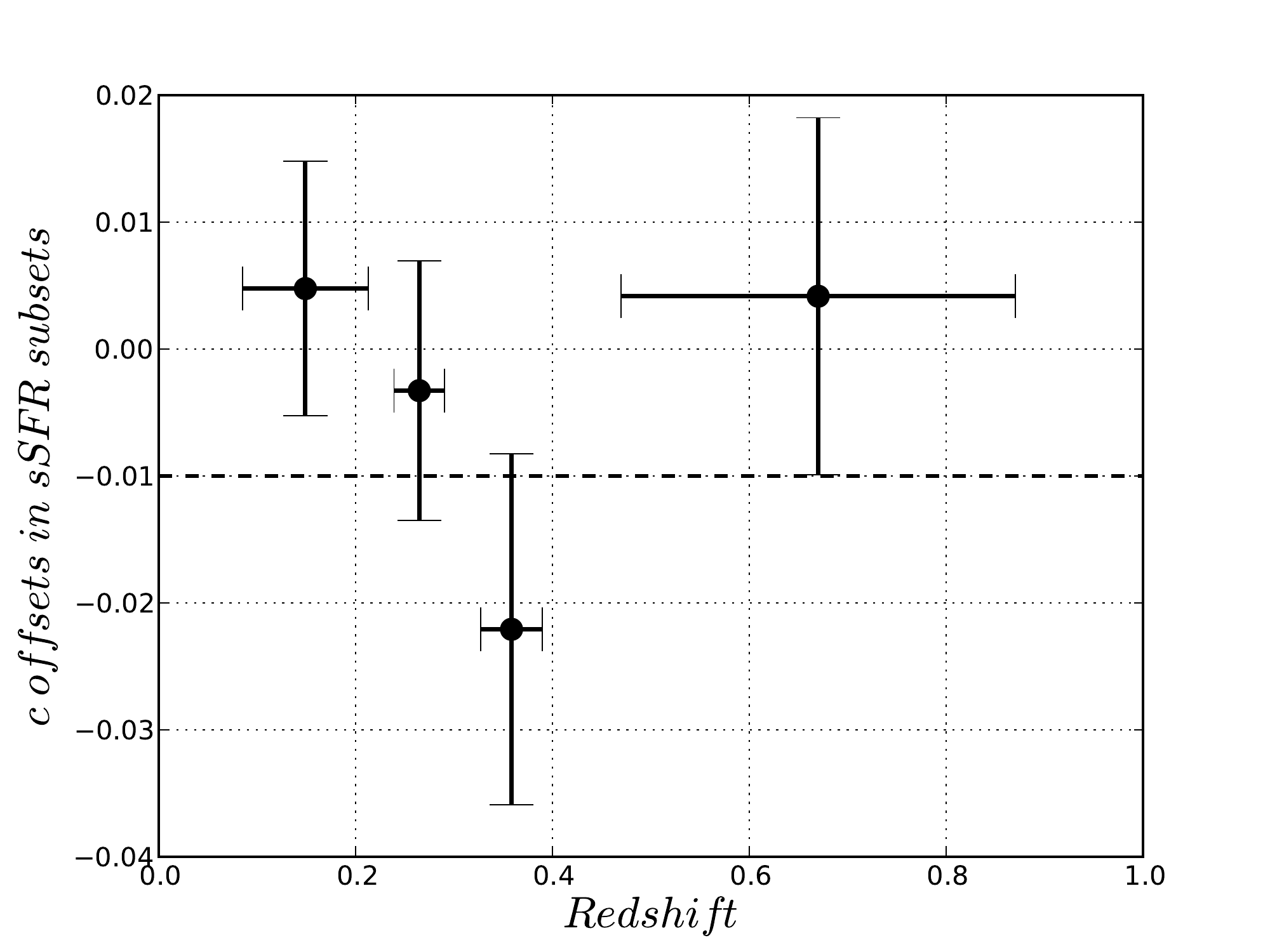}&	
     \includegraphics[width=.33\textwidth]{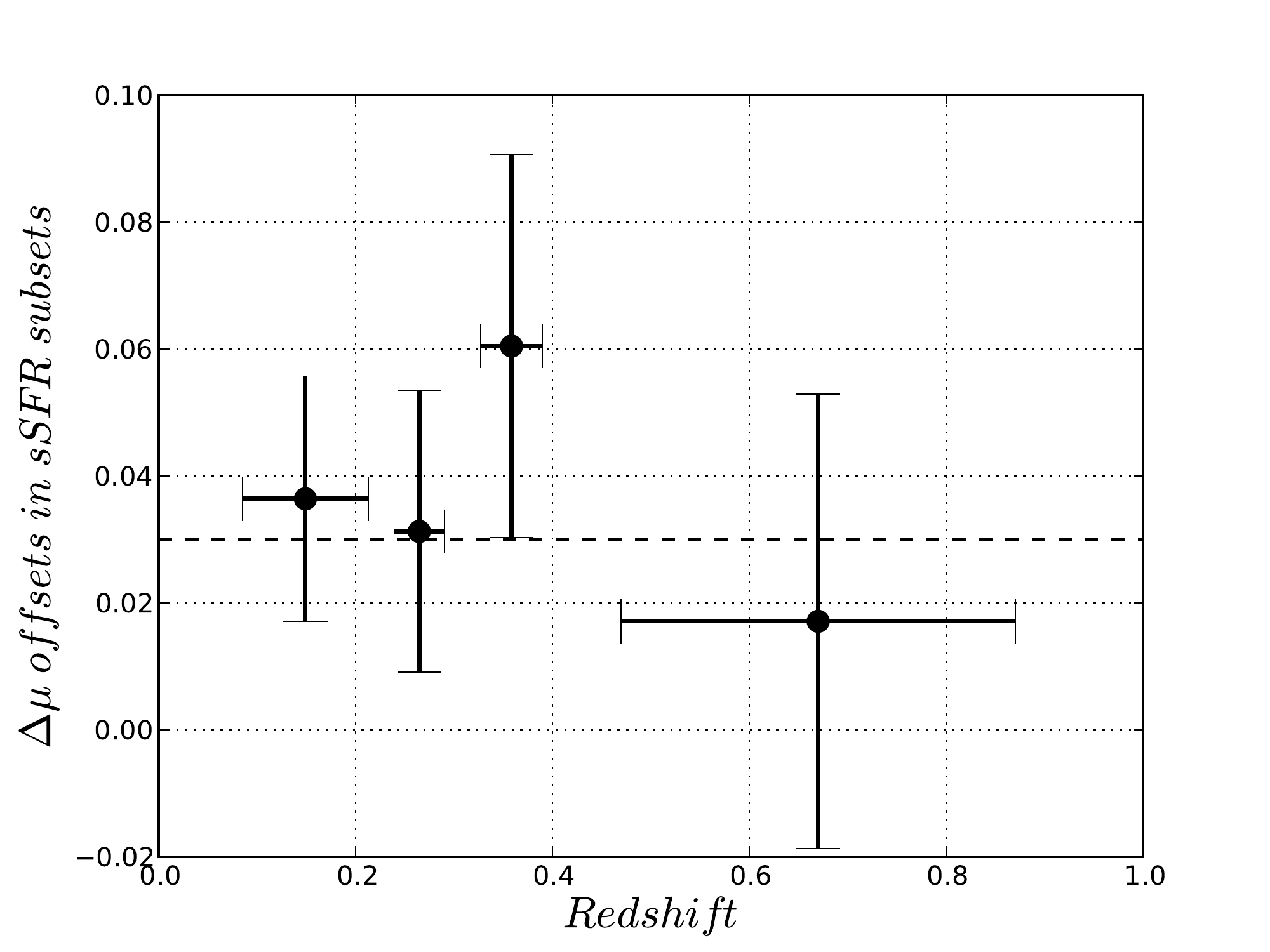}\\
      \includegraphics[width=.33\textwidth]{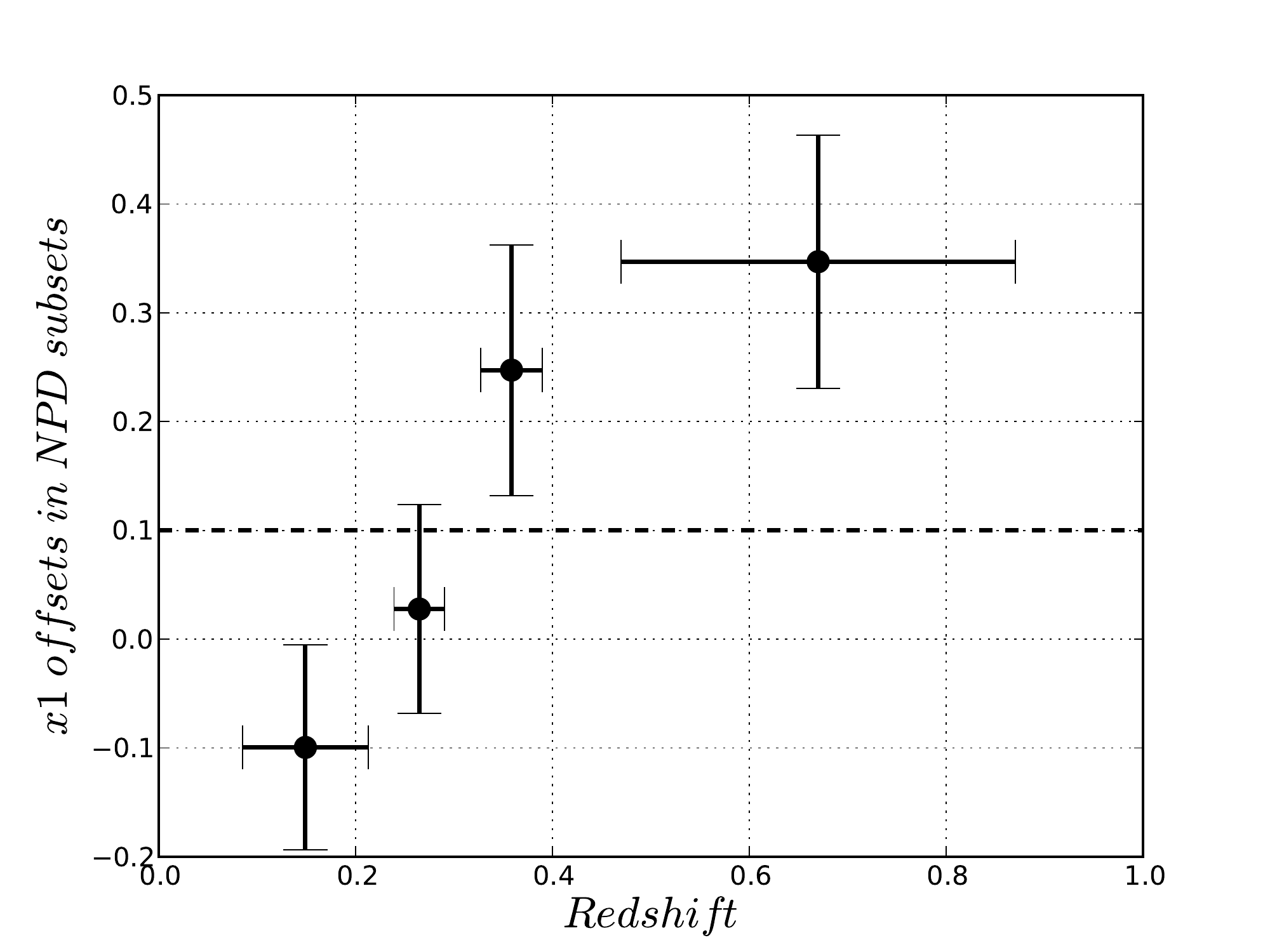}&	              
      \includegraphics[width=.33\textwidth]{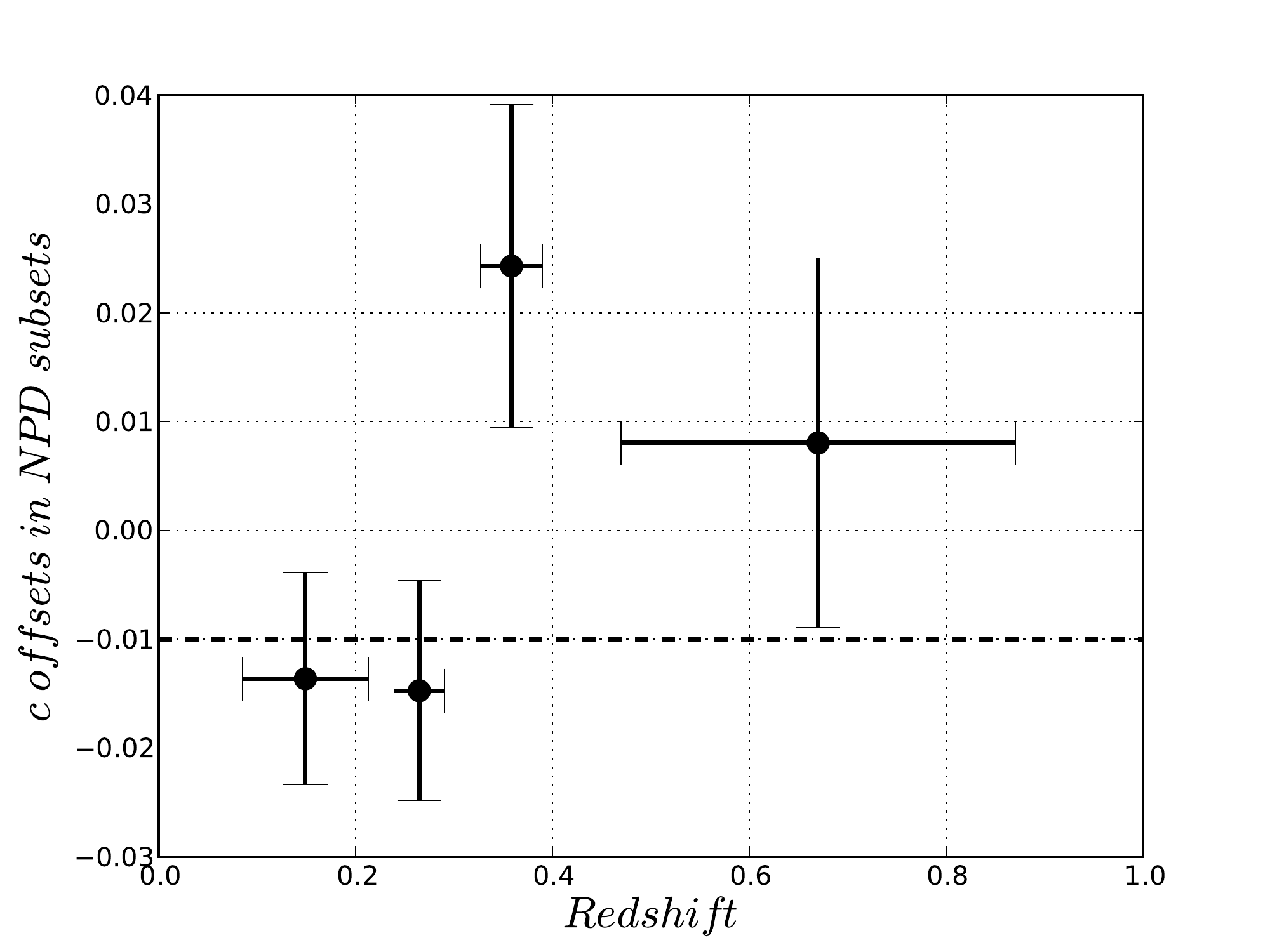}&

      \includegraphics[width=.33\textwidth]{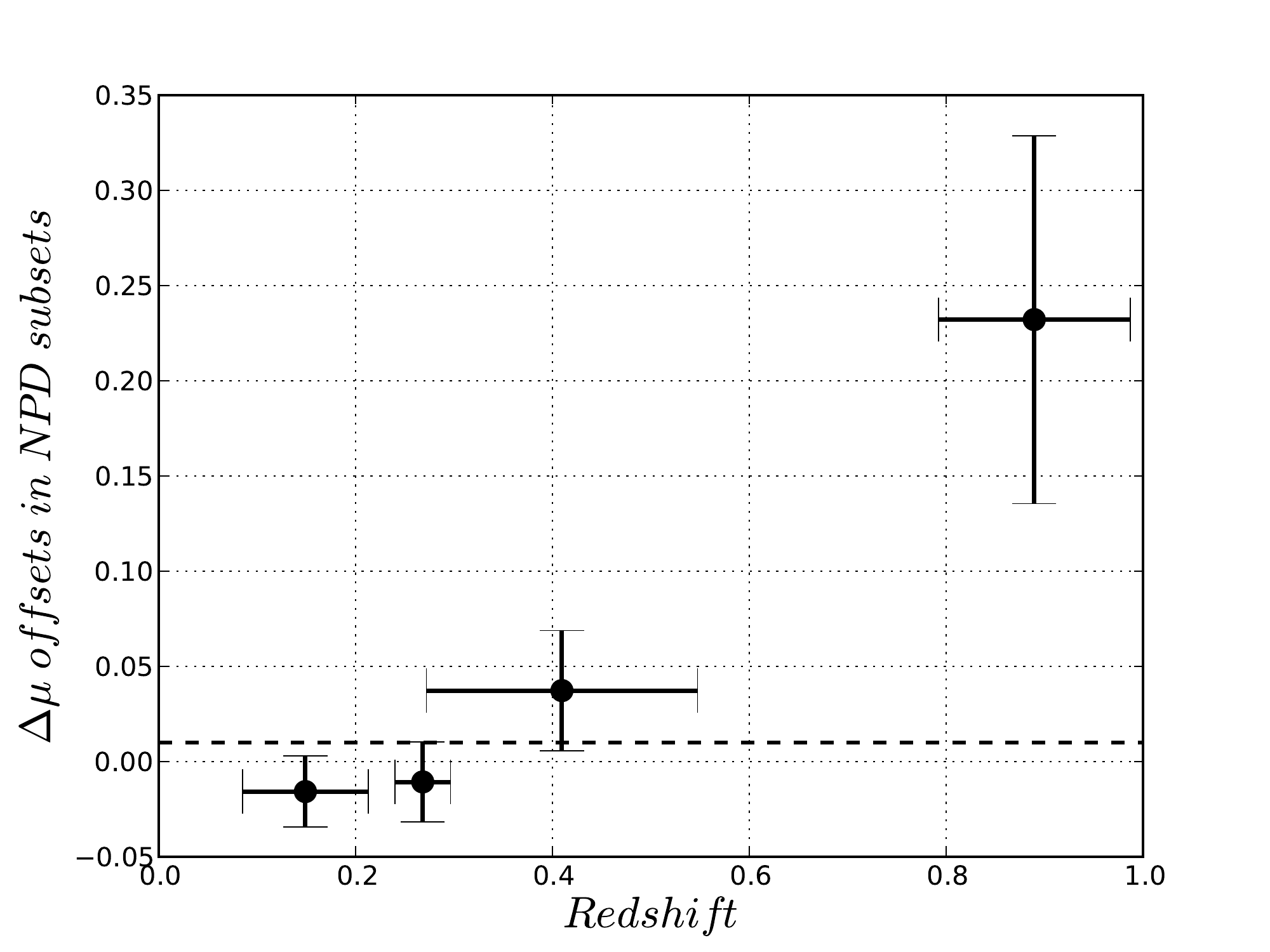}\\
    \includegraphics[width=.33\textwidth]{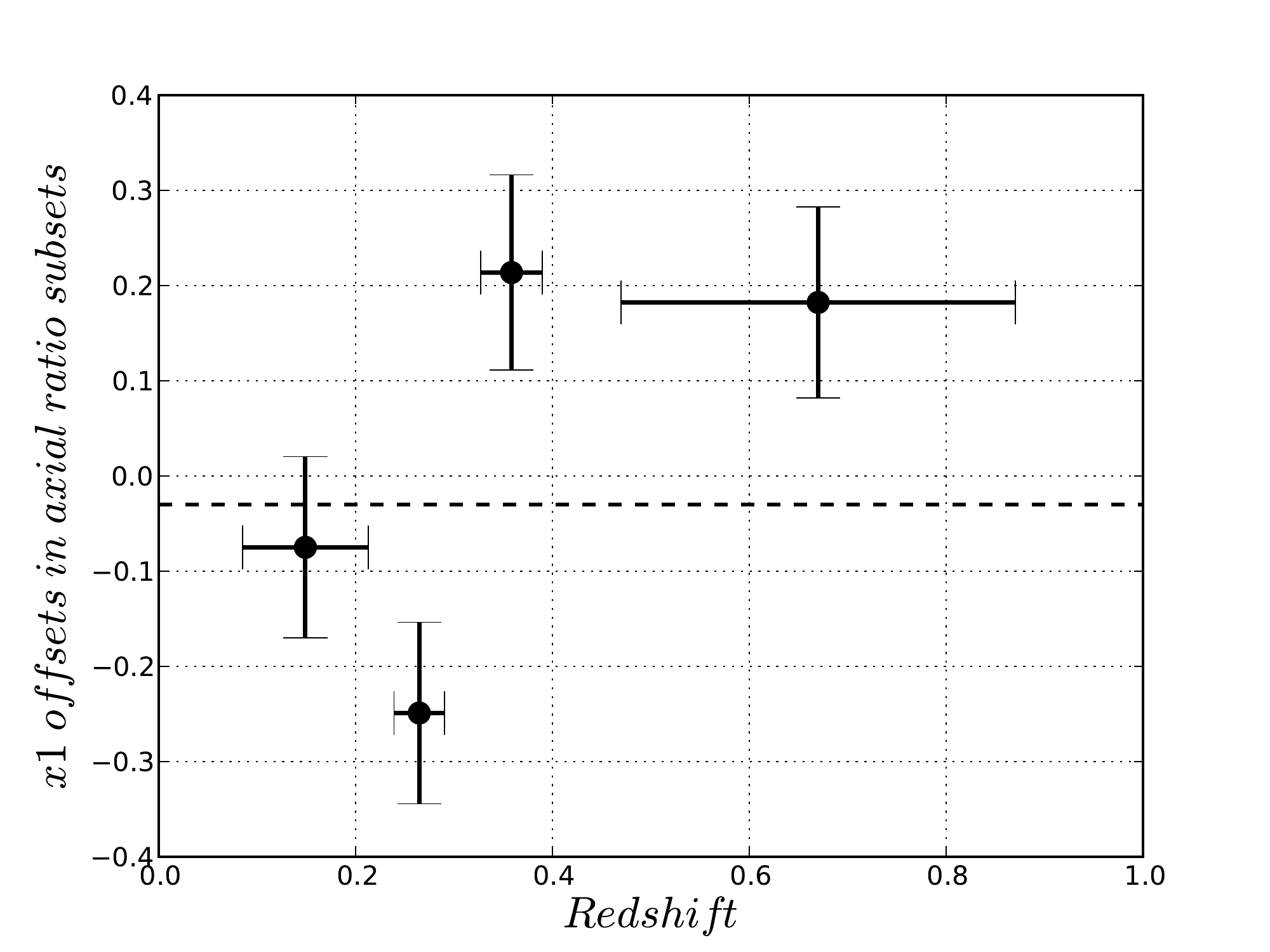}&	
     \includegraphics[width=.33\textwidth]{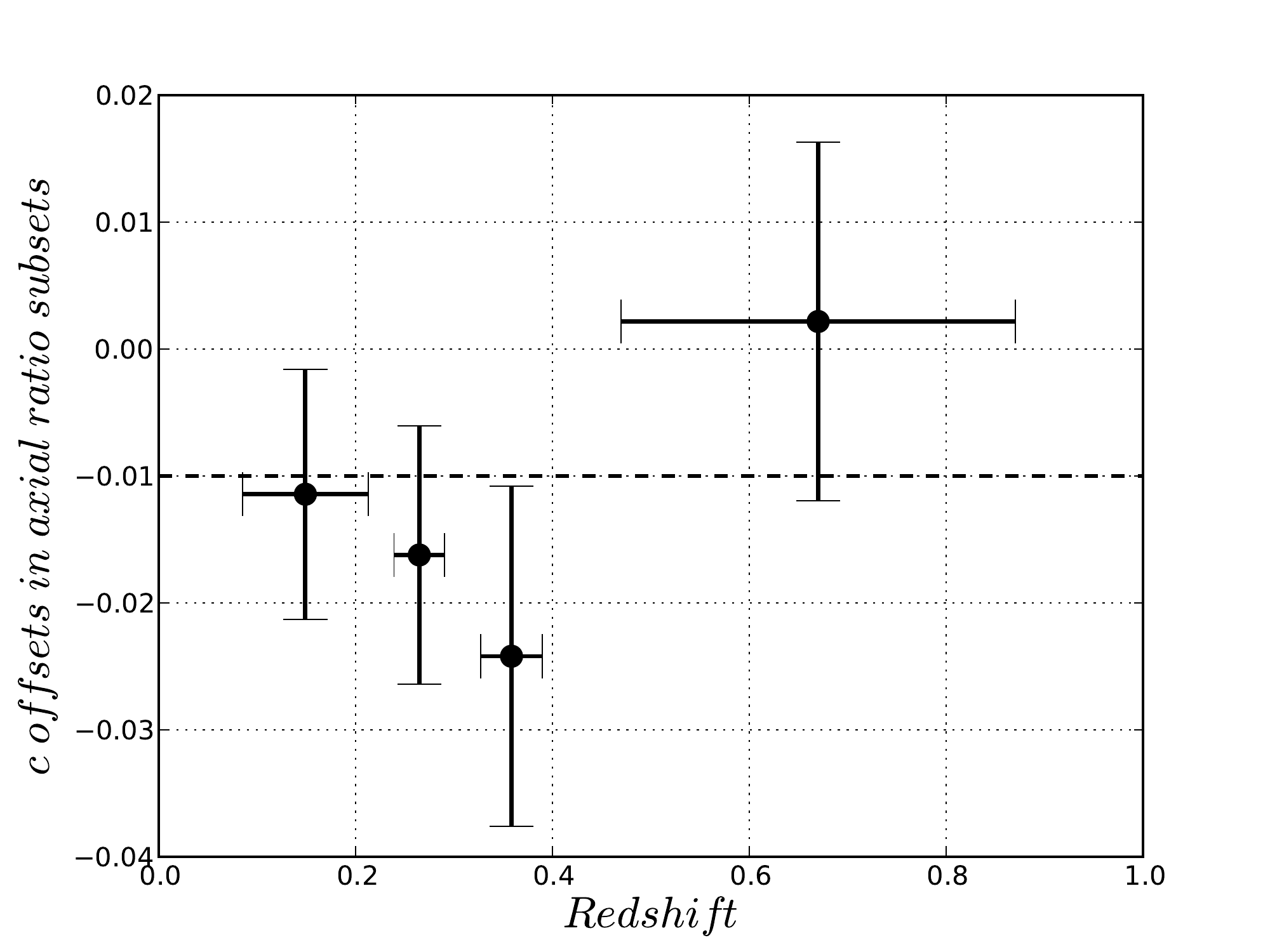}&
      \includegraphics[width=.33\textwidth]{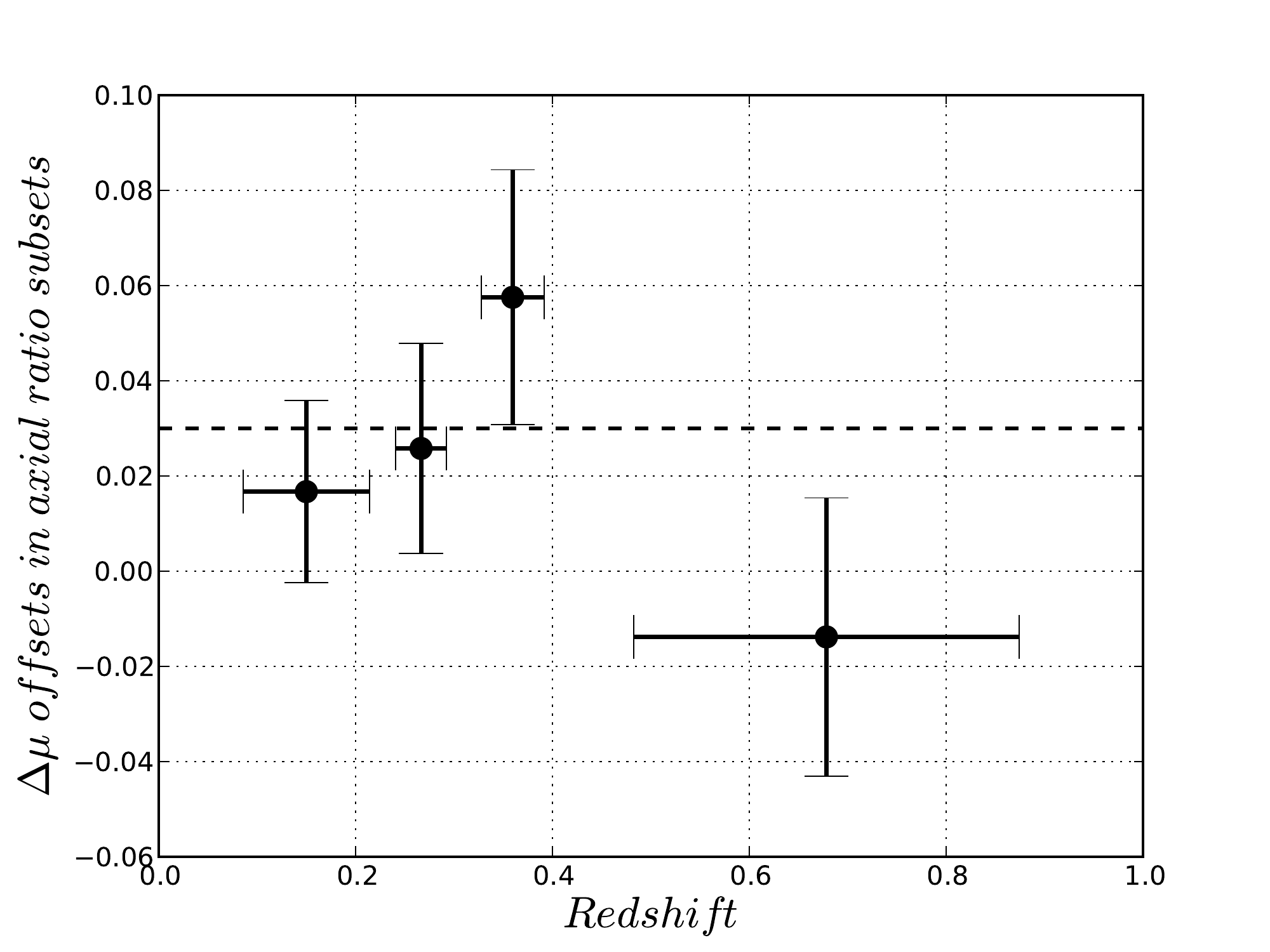}

   \end{tabular} \caption{Redshift evolution of offsets in SN Ia properties between subsets. \emph{Left}: stretch, \emph{middle}: color, and \emph{right}: Hubble residuals. \emph{From top to bottom}: offsets between SN Ia subsets in mass, sSFR, NPD, and axial ratio. In each case dashed lines show the redshift-independent offsets (values from Table~\ref{diffcomb}).}
\label{zall}  
\end{figure*}

We divide our sample into four redshift bins in such a way that each bin contains equal numbers ($\approx$ 334) of SNe Ia. The resulting redshift bins are $\rm z<0.152, \ 0.152\le z< 0.270, \ 0.270\le z \le 0.427, \ and \ z>0.427$. In Fig.~\ref{zall} we show how the offsets evolve with redshift. 

First we see how the offset in stretch changes with redshift. The left panels of Fig.~\ref{zall} shows the redshift evolution of the offset in stretch according to different host properties. The offset in stretch between SNe Ia in low and high mass galaxies does not seem to evolve with redshift. This is also true for the offset in stretch between low and high sSFR hosts. The offset in stretch between the NPD subsets and axial ratio subsets are also consistent with no evolution. 

Next we investigate how the offset in SN Ia color changes with redshift. With host stellar mass, sSFR, NPD, and axial ratio subsets, the offsets does not evolve with redshift (Fig.~\ref{zall}, middle panels). 


Finally, we study how the offset in the Hubble residual changes with redshift (Fig.~\ref{zall}, right panels). In the literature \citep{childress14} the offset in the Hubble residual with galaxy mass is referred to as the Hubble residual mass-step. There are three recent papers where the evolution in the Hubble residual mass-step has been studied. These are \citet{rigault13}, B14, and \citet{childress14}.  \cite{rigault13} give a toy model that predicts evolution in the Hubble residual mass step with redshift. \cite{childress14} predict a change of Hubble residual mass-step utilizing the fact that the spread in galaxy ages decreases with redshift. Our results do not show any significant evolution in the Hubble residual mass-step with redshift. Previously, B14 also obtained a similar conclusion. We do not find any significant evolution of the offset in the Hubble residual with host sSFR, axial ratio, and NPD.

\subsection{Shifts in Cosmological Parameters Between Subsets}\label{ow}
We have seen that SN Ia luminosities (Hubble residuals) vary mostly with host galaxy mass. In a given sample, where SNe Ia are not separated in terms of host mass, this luminosity variation will remain. It is natural to investigate whether this biases the cosmological inference or not. A way to investigate this is to obtain the cosmological inference from the subsets and calculate the shifts between a pair of subsets. The subsets that we fit are listed in Table \ref{diffcomb}. We first describe the cosmology fitting method following \cite{uddin17a} and B14. Thereafter we present our results.

 We compute observed distance modulus using derived $B$-band peak magnitude ($m_B$), light-curve width ($x_1$), and color($c$) of SNe Ia, and the fitted slopes of the color-luminosity ($\beta$) and  light-curve width-luminosity relations ($\alpha$). The slopes $\alpha$ and $\beta$ and the absolute magnitude $M_B$ come from the cosmological fit. We compute observed distance moduli as:

\begin{equation}
\mu_o = m_B + \alpha x_1 -\beta c -M_B 
\end{equation}

The cosmology dependence in the above equation resides in luminosity distance $d_L$. We define the Hubble residual as $\Delta \mu \equiv \mu_o-\mu_m$. Here $\mu_m$ is the model distance moduli. Hubble residuals give us a measure of how much brighter or dimmer SNe Ia are with respect to the best fit cosmology. We fit cosmological parameters ($\Omega_m$, $w$) together ($w$CDM) or $\Omega_m$ alone ($\Lambda$CDM)  along with nuisance variables ($\alpha$, $\beta$, $M_B$). We minimize:

\begin{equation}
\bf{\chi^2 = \Delta \mu^{\dagger}C^{-1} \Delta \mu}
\label{chi}
\end{equation}

Following B14, $\bf{C}$ is defined as
\small

\begin{equation}
\bf C = AC_{\eta}A^{\dagger} + diag \bigg(\frac{5\sigma_z}{zlog10}\bigg)^2 + diag(\sigma^2_{lens}) + diag(\sigma^2_{int})
\label{covar}
\end{equation}

\normalsize
where $\bf{C_{\eta}}$ contains statistical and systematic uncertainties. Matrix $\bf A$ is defined in equation 12 in B14. The second term of Equation \ref{covar} is due peculiar velocities and can be expressed as $c\sigma_z=\mathrm{150\ kms^{-1}}$ (\citealt{conley11}). The third term is due gravitational lensing, which is expressed as $\sigma_{lens}=0.055\times z$ \citep{jonsson10}.

The last term in Equation \ref{covar} represents the intrinsic dispersion $\sigma_{int}$,  which represents the excess dispersion after standardization. We do know this term is needed as SNe Ia are not fully standardized and we yet do not know what causes the extra dispersion in their luminosities (see \citealt{conley11}). For the global sample, we first fit $\sigma_{int}$ as a free parameter using the restricted log-likelihood method (Equation 14 of B14) and derive the best-fit value for $\sigma_{int}$. We use this value to derive final cosmological parameters by minimizing Equation \ref{chi}. 

In our analysis, we use Malmquist bias corrections. For the spectroscopically confirmed sample, we have used the corrections listed in B14, originally calculated in \citep{conley11}. Malmquist bias corrections were calculated for each survey and they are redshift dependent. For the photometrically classified sample, we have calculated redshift dependent Malmquist bias using the trend found in B14. 

We use Markov Chain Monte Carlo (MCMC) method to explore the $\chi^2$ likelihood for the Equation \ref{chi} and derive best-fit cosmological parameters. We then take a pair of subsets based on the host property and calculate the shifts between the cosmological parameters. Table \ref{disp} shows these shifts for both $\Lambda$CDM and $w$CDM cosmological frameworks. It is easy to see that shifts in $\Omega_M$ and $w$ are insignificant.

The importance of our finding is that, even though SN Ia properties correlate with some of the properties of hosts, they do not affect the cosmological inference. \cite{sullivan11} have derived cosmological parameters with SNe Ia from SNLS and low-redshift surveys and have found that cosmological parameters remain consistent between two samples segregated by host galaxy mass (see Table 8 of that paper).  Our results support this earlier finding with a larger SNe Ia sample. While this is true for the current sample, we may start to see difference in future SN Ia samples, which will be even larger than the one used here.

\begin{table*}[htp]

\caption{Shifts in cosmological parameters between pairs of subsets. Significances of shifts are shown in parentheses.}
\begin{center}
\begin{tabular}{lccc}
\hline
& $\Lambda$CDM & $w$CDM\\
SN Ia Subsets & $\delta \Omega_m$ &  $\delta \Omega_m$ & $\delta w$ \\
\hline
Low-high mass hosts & $0.06 \ ( 0.5\sigma)$& $0.12 \ (0.8\sigma)$ & $0.15 \ (0.2\sigma)$\\
Low-high sSFR hosts & $0.02\ (0.2\sigma)$& $0.13 \ (0.7\sigma)$& $0.23 \ (0.5\sigma)$ \\
Low-high NPD hosts& $0.05 \ (0.4\sigma)$& $0.16 \ (0.6\sigma)$ & $0.22 \ (0.3\sigma)$\\
Low-high axial ratio hosts& $0.01 \ (0.1\sigma)$& $0.13 \ (0.7\sigma)$& $0.39 \ (0.8\sigma)$\\
 
\hline
\end{tabular}
\end{center}
\label{disp}
\end{table*}%

\subsection{Intrinsic Scatter ($\sigma_{int}$)}\label{in}

Here we examine the homogeneity of SN Ia luminosities in various subsets. To do this, we set the intrinsic scatter ($\sigma_{int}$) as a free parameter in the fit. We show them in Table \ref{int_scat}. For SNe Ia in hosts with high-sSFRs, we obtain $\sigma_{int}=0.08\pm 0.01$ mag. The full sample has $\sigma_{int}=0.12\pm 0.01$ mag. Therefore, we observe a $4\sigma$ reduction in the intrinsic scatter. This is an indication that SNe Ia in high-sSFR hosts are better standard candles. With more data from ongoing and future surveys, the significance of this result might increase further. This result is expected according to the prediction made in \cite{childress14}, which says that SNe Ia are more homogeneous in high star-forming hosts due to smaller delay times.

In the study by \cite{rigault13}, SNe Ia associated with local $H\alpha$ emission, which is related to active star-formation, are found to be more homogeneous compared to the whole sample. Recently \citep{kelly15} have found that SNe Ia exploding in regions of high star formation surface densities have $\sigma_{int}\approx 0.065-0.075$ mag. We note that these two studies have been performed with a handful of SNe Ia and are restricted to low-redshift ($z<0.05$). Our finding is therefore important since it covers a wider redshift range and contains more SNe Ia.

 \begin{table*}[htp]
\caption{Intrinsic scatter ($\sigma_{int}$) in different subsets. The least scatter is found for SNe Ia in hosts with high-sSFRs. Uncertainties in each case is $\sim 0.01$. }
\begin{center}
\begin{tabular}{cccccccc}\\
\hline
Low-mass & High-mass & Low-sSFR & High-sSFR & Low-NPD & High-NPD & Low-axial & High-axial\\
\hline
0.092 & 0.137& 0.148 & 0.088 & 0.124 & 0.117& 0.110& 0.132\\
\hline
\end{tabular}
\end{center}
\label{int_scat}
\end{table*}%

\section{Discussion}\label{dis}
\subsection{Comparing Results with Previous Studies}
Our results agree with previous studies qualitatively. Using a low-redshift sample, \cite{kelly10} found a Hubble residual offset of $0.094\pm 0.045$ mag between low and high mass galaxies. With a larger sample, \cite{sullivan10} also found that SNe Ia are brighter in massive hosts by 0.08 mag with 4$\sigma$ confidence. Later studies such as \cite{childress13} and  \cite{pan14} also had similar findings. Our finding that SNe Ia are bright in massive hosts by 0.05 mag with $5.3\sigma$ confidence is in agreement with other studies. Also our finding that SNe Ia have narrower stretches in massive hosts agrees with previous findings such as \cite{sullivan10} and  \cite{pan14}. 

When comparing the Hubble residual difference in low- and high-sSFR hosts, \cite{pan14} found a difference of 0.070 mag (1.7$\sigma$) in contrast to our value which is 0.04 mag (2.1$\sigma$). Previously \cite{dandrea11}, who used spectra to derive the sSFR, found a 3.1$\sigma$ difference where sSFR rate is derived from host spectra. 

For SDSS-II SNe Ia, \cite{galbany12} separated hosts morphologically and found that SN Ia color decreases significantly (4$\sigma$) with the projected distance from host centers for spiral galaxies and that light-curve widths decrease with the projected distance for elliptical galaxies. They found no significant correlation between the Hubble residual and the projected distance. We have not performed a morphological classification as that is possible only with low-redshift hosts (e.g. not possible for most of the SNLS SNe Ia). For our sample, we find that SNe Ia that are further away from the centers of the hosts have lower stretches by 0.1 (2.1$\sigma$), are slightly bluer by 0.01 (0.4$\sigma$), and are fainter by 0.01 mag (0.3$\sigma$). The variation of SN Ia properties and Hubble residual with host galaxy axial ratio has not been studied elsewhere and therefore we cannot compare our results with the literature.

\subsection{Use of a Different SED Fitting Code}\label{lephare}

As we have discussed in Section \ref{hostprop}, galaxies are found to form diagonal bands along the mass-SFR plane (see Fig.~\ref{gmain}). The amount of banding can be reduced if we use a different SED fitting code, such as Le PHARE (\citealt{lephare}). We repeated part of the analysis using the masses and SFRs from Le PHARE. Our results did not change significantly.

\section{Summary and Outlook}\label{summary}

In this paper we have compiled a sample of 1338 spectroscopically confirmed and photometrically classified SNe Ia from the CSP, CfA, SDSS-II, and SNLS surveys and have uniformly derived the properties of SNe Ia and their hosts. We have studied the offsets in SN Ia properties between subsets. We have also examined the evolution of these offsets with redshift. We have fitted cosmological models to each subset. Here we summarize our main findings:

\begin{enumerate}[label=(\alph*)]
\item SNe Ia are significantly more luminous in high-mass galaxies ($\sim 5.0 \sigma$).

\item Faster declining SNe Ia are preferentially found in high mass galaxies ($9.2\sigma$) and in galaxies with low sSFR ($5.1\sigma$).

\item These differences are independent of redshift.

\item Shifts in cosmological parameters in both $\Lambda$CDM and in $w$CDM frameworks are insignificant between subsets.

\item SNe Ia that explode in high-sSFR hosts have the least intrinsic scatter in their light-curve corrected luminosities ($\rm \sigma_{int}=0.08\pm 0.01 \  mag$)

\end{enumerate}

Several studies have tried to explain the variation of SN Ia properties with their host. For example, radioactive decay of $\rm ^{56}Ni$ is thought to power the observed luminosity and stretch of SNe Ia. \cite{timmes03} suggested a relationship between ejected $\rm ^{56}Ni$ mass and host metallicity. If such a relationship exists, then it might account for some of these correlations that we see in this work because of the galaxy mass-metallicity relation.

To better constrain host galaxy properties, it will be useful to add additional information from other parts of the electromagnetic spectrum. Along with optical photometry, data from GALEX (ultraviolet) and Spitzer (near and mid infra-red) can be added to improve (\citealt{gupta11}) the measurements of host galaxy properties. High-resolution measurements of the local environments near to the SNe Ia explosion sites, using optical integral field spectroscopy and high-resolution imaging, may give more insight onto the systematics in SNe Ia cosmology coming from host environments.

Current and future (e.g., DES, LSST respectively) surveys will discover many thousands SNe using photometry alone. Only a small fraction will have spectroscopic confirmation\footnote{For example, about 20$\%$ of the DES SNe will have spectroscopic confirmation. }. For DES, redshifts for most SNe will be obtained from host spectra while for LSST, most of the redshifts will come from SED fitting (photo-z). DES alone will produce $\sim 3000$ SNe Ia in its five year campaign \citep{bernstein12}. A great advantage of having a large number of SNe Ia from a single survey is better control of systematic uncertainties. The challenge that will remain is to reduce astrophysical systematic errors due to SN Ia-host correlations. Our methodology that we have established in this work can be applied to DES SNe Ia. It may eventually lead to tighter constraints on the properties of dark energy. We discuss cosmological constraints from a DES-like SNe Ia sample in a separate paper (\citealt{uddin17a}).

\acknowledgments 

\footnotesize{We thank Chris Blake of Swinburne University of Technology for helpful discussion regarding statistical analysis. We also thank Mark Sullivan of University of Southampton for valuable comments on the paper. Part of this research was conducted by the Australian Research Council Centre of Excellence for All-sky Astrophysics (CAASTRO), through project number CE110001020. Syed A Uddin was supported by the Chinese Academy of Sciences President's International Fellowship Initiative Grant No. 2016PM014. This research made use of Astropy, a community-developed core Python package for Astronomy (\citealt{astropy13}). }

\bibliography{bibSyedUddin}





\begin{landscape}

\appendix
\normalsize
\section{SALT2.4 Light-Curve Parameters}\label{lcdata}
Here we present SALT2.4 light-curve fit parameters (i.e. colour, stretch and peak $B$-band magnitude) for the SN Ia sample. In Table \ref{lcdef}, we describe each entry. In Table \ref{lcsn} we present the light-curve fit parameters.

\begin{table}[htp]
\caption{Description of the entries in Table \ref{lcsn}.}
\begin{center}
\begin{tabular}{ll}
\hline
Item & Description\\
\hline
Name & Name of the SN Ia\\
$zhel$ & SN Ia heliocentric redshift\\
$zcmb$ & SN Ia CMB reference redshift\\
$m_B$ & SN Ia peak B-band apparent magnitude\\
$em_B$ & Error in $m_B$\\
$x_1$ & SN Ia Stretch or light-curve width\\
$ex_1$ & Error in $x_1$\\
$c$ & SN Ia colour \\
$ec$ &Error in $c$\\
$cov(m_Bx_1)$ & Covariance between $m_B$ and $x_1$ (in units of $10^{-4}$)\\
$cov(m_Bc)$ & Covariance between $m_B$ and $c$ (in units of $10^{-4}$)\\
$cov(x_1c)$ & Covariance between $x_1$ and $c$ (in units of $10^{-4}$)\\
Source & Source of the SN Ia\\
Type & Type of SN Ia (s = spectroscopically confirmed; p = photometrically classified)\\
\hline 
\end{tabular}
\end{center}
\label{lcdef}
\end{table}%

\scriptsize
\begin{longtable}{@{\extracolsep{\fill}}lcccccccccccll@{}}

\caption{SALT2.4 light-curve properties for spectroscopically confirmed SNe Ia. Full table is available online.}\\
\hline
SN Name & $zhel$ & $zcmb$&$m_B$ & $em_B$ & $x_1$ & $ex_1$ & $c$ & $ec$ & $cov(m_Bx_1)$ & $cov(m_Bc)$ & $cov(x_1c)$ & Source & Type\\
\hline
\endfirsthead
$\ldots$ continued from last page\\
\hline
SN Name & $zhel$ & $zcmb$&$m_B$ & $em_B$ & $x_1$ & $ex_1$ & $c$ & $ec$ & $cov(m_Bx_1)$ & $cov(m_Bc)$ & $cov(x_1c)$ & Source & Type\\
\hline
\endhead
\hline
SNLS05D3jr&0.370&0.370&22.629&0.087&-0.893&0.121&0.049&0.023&-0.261&0.489&-0.197&SNLS&s\\
SNLS05D3jq&0.579&0.578&23.295&0.089&1.489&0.208&0.002&0.029&0.453&0.435&-0.168&SNLS&s\\
SDSS15776&0.317&0.318&21.847&0.122&-2.282&0.561&-0.146&0.050&16.160&0.877&6.370&SDSS&s\\
SDSS6057&0.067&0.067&18.640&0.113&-0.421&0.108&0.108&0.025&0.458&0.525&-0.004&SDSS&s\\
SNLS05D3jb&0.745&0.745&23.920&0.095&0.939&0.232&-0.063&0.052&3.101&0.155&-0.043&SNLS&s\\
SDSS16072&0.286&0.285&21.533&0.127&0.276&0.625&-0.016&0.054&30.689&2.286&13.875&SDSS&s\\
SDSS16073&0.155&0.155&20.257&0.112&0.723&0.185&-0.007&0.024&1.746&0.569&0.895&SDSS&s\\
SNLS04D2cf&0.369&0.370&22.463&0.096&-0.862&0.218&-0.014&0.035&-5.152&1.364&-2.585&SNLS&s\\
SNLS05D3jk&0.736&0.736&23.692&0.094&0.774&0.180&-0.132&0.048&1.352&0.268&-0.342&SNLS&s\\
SNLS05D3jh&0.718&0.718&23.724&0.095&-0.614&0.184&-0.138&0.052&2.357&0.408&0.437&SNLS&s\\
SNLS06D2bk&0.499&0.500&23.277&0.089&0.471&0.277&0.005&0.032&3.618&0.478&1.062&SNLS&s\\
sn2006te&0.032&0.032&16.531&0.038&-0.060&0.100&-0.072&0.030&-1.235&0.977&-0.862&CfA3&s\\
\hline
\label{lcsn}
\end{longtable}

\newpage
\normalsize

\section{Host Galaxy Properties}\label{hostdata}
Here we present host galaxy photometry and physical properties. In Table \ref{hostdef}, we describe each entry. In Table \ref{hosts}, we present host galaxy photometry and physical properties.

\begin{table}[htp]
\caption{Description of the entries in Table \ref{hosts}.}
\begin{center}
\begin{adjustbox}{width=1\textwidth}
\begin{tabular}{ll}
\hline
Item & Description\\
\hline
Name & Name of the SN Ia\\
$\rm Log \  M_{stellar} \ (M_{\odot})$ & Host galaxy stellar mass in solar units\\
$\rm  Log \ sSFR \ (yr^{-1})$ & Host galaxy specific SFR\\
$\rm NPD$ & Normalized projected distance between SN Ia and host galaxy centres\\
$\rm Axial \  Ratio$ & Ratio between the semi-major and semi-minor axes of the host galaxy\\
$\rm RA$ & Right Ascension of host galaxy\\
$\rm DEC$ & Declination of host galaxy\\
\hline
\end{tabular}
\end{adjustbox}
\end{center}
\label{hostdef}
\end{table}%

\scriptsize
\setlength\LTleft{0pt}
\setlength\LTright{0pt}
\begin{longtable}{@{\extracolsep{\fill}}lcccccc@{}}
\caption{Host galaxy photometry and properties. Full table is available online. }\\
\hline
SN Name &$Log \  M_{stellar} \ (M_{\odot})$ & $Log \ sSFR \ (yr^{-1})$ & $NPD$ & $Axial \  Ratio$ &$RA$&$DEC$\\
\hline
\endfirsthead
$\ldots$ continued from last page\\
\hline
SN Name & $Log \  M_{stellar} \ (M_{\odot})$ & $Log \ sSFR \ (yr^{-1})$ & $NPD$ & $Axial \  Ratio$ &$RA$&$DEC$\\
\hline
\endhead
\hline
SNLS05D3jr&9.976&-9.963&1.326&0.607&214.86978&52.86485\\
SNLS05D3jq&10.355&-9.520&1.424&0.947&215.43933&53.02992\\
SDSS15776&11.061&-99.000&2.810&0.923&32.82944&-0.99826\\
SDSS6057&10.108&-9.560&0.162&0.862&52.55360&-0.97459\\
SNLS05D3jb&10.669&-99.000&1.330&0.555&215.54166&52.87813\\
SDSS16072&10.874&-99.000&1.575&0.862&3.12444&-0.97731\\
SDSS16073&9.468&-8.952&0.350&0.713&8.10776&-1.05400\\
SNLS04D2cf&11.068&-10.660&1.459&0.726&150.48370&1.87948\\
SNLS05D3jk&10.008&-9.504&0.272&0.577&214.19772&52.59248\\
SNLS05D3jh&11.254&-9.554&0.394&0.778&214.35577&52.61877\\
SNLS06D2bk&9.041&-9.232&0.864&0.915&149.67863&2.17198\\
sn2006te&9.159&-9.000&0.919&0.576&122.92901&41.55467\\
\hline
\label{hosts}
\end{longtable}%
\end{landscape}

\end{document}